\documentclass[useAMS,usenatbib]{mn2e}

\pdfoutput=1
\usepackage{graphicx} 
\usepackage{epsfig}   
\usepackage{amsmath}
\usepackage{amssymb} 
\usepackage{mathrsfs}
\usepackage{url}
\usepackage{txfonts}
\usepackage{natbib}
\usepackage{units}
\usepackage{longtable, lscape, multirow}
\title[Fundamental parameters of 16 late-types stars]{Fundamental parameters of 16 late-type stars derived from their angular diameter measured with VLTI/AMBER\thanks{Based on observations made with ESO telescopes at the Paranal Observatory under Belgian VISA Guaranteed Time programme ID 083.D-029(A/B), 084.D-0131(A/B), 086.D-0067(A/B/C) } }
\author[P. Cruzal\`ebes et al.]{P. Cruzal\`ebes,$^{1}$\thanks{E-mail:
pierre.cruzalebes@oca.eu} A. Jorissen,$^{2}$ Y. Rabbia,$^{1}$ S. Sacuto,$^{3,4}$ A. Chiavassa,$^{1,2}$  E. Pasquato,$^{2}$ \newauthor 
B. Plez$^{5}$, K. Eriksson$^{3}$, A. Spang,$^{1}$ and O. Chesneau$^{1}$ \\
$^{1}$Laboratoire Lagrange, UMR 7293, Universit\'e de Nice-Sophia Antipolis, CNRS, Observatoire de la C\^ote d'Azur, Bd de l'Observatoire, B.P. 4229, \\
F-06304 Nice cedex 4, France\\
$^{2}$Institut d'Astronomie et d'Astrophysique, Universit\'e Libre de Bruxelles, Campus Plaine C.P. 226, Bd du Triomphe, B-1050 Bruxelles, Belgium\\
$^{3}$Department of Astronomy and Space Physics, Uppsala Astronomical Observatory, Box 515, S-751 20 Uppsala, Sweden\\
$^{4}$Institute of Astronomy, University of Vienna, T\"urkenschanzstra\ss e 17, A-1180 Vienna, Austria\\
$^{5}$Laboratoire Univers et Particules, Universit\'e Montpellier II, CNRS, F-34095 Montpellier cedex 5, France
}

\begin{document}

\date{Version 2013 May 27}

\pagerange{\pageref{firstpage}--\pageref{lastpage}} \pubyear{tbd}

\maketitle

\label{firstpage}

\begin{abstract}
  {Thanks to their large angular dimension and brightness, red giants and supergiants are privileged
targets for optical long-baseline interferometers. Sixteen red giants and supergiants have been
observed with the VLTI/AMBER facility over a two-years period, at medium spectral resolution
(${\mathscr{R}=1500}$) in the $K$ band. 
The limb-darkened angular diameters are derived from fits of stellar atmospheric models on 
the visibility and the triple product data. 
The angular diameters do not show any significant temporal variation, except for one target: TX~Psc, which shows a variation of 4\% using visibility data. For the eight targets previously measured by Long-Baseline Interferometry (LBI) in the same spectral range,
the difference between our diameters and the literature values is
less than 5\%, except for TX~Psc, which shows a difference of 11\%. For the 8 other targets, the present angular diameters
are the first measured from LBI. Angular diameters are then used 
to determine several fundamental stellar parameters, and to locate these
targets in the Hertzsprung-Russell Diagram (HRD). Except for the enigmatic Tc-poor low-mass carbon star W~Ori, the location of Tc-rich stars
in the HRD matches remarkably well the thermally-pulsating AGB, as it is predicted by the stellar-evolution models. 
For pulsating stars with periods available, we compute the pulsation constant and locate the stars along the
various sequences in the Period -- Luminosity diagram. 
We confirm the increase in mass along the pulsation sequences, as predicted by the theory, 
except for W~Ori which, despite being less massive, 
appears to have a longer period than T~Cet along the first-overtone sequence. 
}
\end{abstract}


\section{Introduction} \label{SecIntro}

The direct measurement of stellar angular diameters has been the principal goal of most attempts with astronomical interferometers since the pioneering work of \citet{michelson21}. For stars of known distance, the angular diameter $\phi$, combined with the parallax $\varpi$, yields the stellar radius ${\mathcal{R}=0.5{\phi}/{\varpi}}$, where $R$ is in AU. When combined with the emergent flux at the stellar surface, linked to the effective temperature $T_{\mathrm{eff}}$, the stellar radius $\mathcal{R}$ leads to the absolute luminosity ${\mathcal{L}=4\upi\mathcal{R}^2\sigma T_\mathrm{eff}^4}$, where $\sigma$ is the Stefan-Boltzmann constant.

These quantities are essential links between the observed properties of stars and the results of theoretical calculations on stellar structure and atmospheres \citep{baschek91,scholz97,dumm98}.

Because of their comparatively large dimension, late-type giants and supergiants are suitable targets for modern Michelson interferometers, reaching accuracies better than a few percent \citep[see e.g.,][]{vbelle96,millan05}. With radii larger than \unit[1]{AU}, many nearby giants subtend relatively large angular diameters ($>$ \unit[20]{mas} at \unit[100]{pc}). They also have high brightnesses in the near infra-red, allowing interferometric measurements with high Signal-to-Noise Ratio (SNR).

Using the ESO/VLTI facility, we initiated in 2009 a long-term program with the ultimate goal of investigating the presence of Surface-Brightness Asymmetries (SBAs), and of their temporal behaviour, following the pioneering work of \citet{ragland06}. This issue is addressed in a companion paper (Cruzal\`ebes et al., submitted to MNRAS). The AMBER instrument is well suited for that purpose, since it provides phase closures at medium spectral resolution in $K$. This goal prompted us to select our targets all over the red-giant and supergiant regions of the HR Diagram (HRD). 
Investigation of SBAs is important in the framework of the Gaia astrometric satellite \citep{perryman01,lindegren08}, since the presence of time-variable SBAs may hinder its ability to derive accurate parallaxes for such stars \citep[see the discussions by][]{bastian05,eriksson07,pasquato11,chiavassa11a}.

In this paper, we present new determinations of the angular diameters of 16 red giants and supergiants, obtained by combining the fits of limb-darkened disk models using two SPectro-Interferometric (SPI) observables:~the visibility amplitude, and the triple product. The visibility is defined as the ratio of the modulus of the coherent to the incoherent flux, and the triple product as the ratio of the bispectrum to the cubed incoherent flux (see \citealt{cruzalebes13} for details). In Sect.~\ref{SecObs}, we describe the measurement technique, and the sample of observed sources; in Sect.~\ref{SecAngdiam}, we describe the model fitting procedure; in Sect.~\ref{SecSensi}, we study the sensitivity of our results with respect to the fundamental parameters of the model:~linear radius, effective temperature, surface gravity, and microturbulence velocity; in Sect.\ref{SecTempoVar}, we study the possible temporal variability of the angular diameter; in Sect.~\ref{SecFinStellarDiam}, we describe the method for deriving the final angular diameter; in Sect.~\ref{SecDiscuss}, we confront our results with those of the literature. 

Then, our stellar radii are used to infer various fundamental stellar characteristics:~(i)~location in the HRD, and masses derived from a comparison with evolutionary tracks (Sect.~\ref{SecHR}); (ii)~luminosity threshold for the occurrence of technetium on the asymptotic giant branch (AGB), since technetium, having no stable isotopes, is a good diagnosis of the s-process of nucleosynthesis (Sect.~\ref{SecTc}); and (iii)~pulsation mode from the location in the Period -- Luminosity (P -- $\mathcal{L}$) diagram (Sect.~\ref{SecPL}). 

The results and graphical outputs presented in the paper were obtained using the modular software suite \textsc{spidast}\footnote{acronym of \textit{SPectro-Interferometric Data Analysis Software Tool}}, created to calibrate and interpret SPI measurements, particularly those obtained with VLTI/AMBER \citep{cruzalebes08,cruzalebes10,cruzalebes13}. 

Throughout the present paper, uncertainties are reported using the concise notation, according to the recommendation of the Joint Committee for Guides in Metrology \citep{jcgm108}. The number between parentheses is the numerical value of the standard uncertainty referred to the associated last digits of the quoted result.

\section{Introducing the observations} \label{SecObs}

\subsection{Selecting the science targets for the programme} \label{SubSecProg}

The sample contains supergiants and long-period variables (LPVs), bright enough (m$_\mathrm{K} < 2$) to be measured by the VLTI subarray ($\unit[1.80]{m}$ auxiliary telescopes) with high SNR. In Table~\ref{TabObsSci}, we compile their relevant observational parameters, including possible multiplicity and variability.  
On one hand, the scientific targets must be resolved well enough, which results in visibilities clearly smaller than unity. On the other hand, visibilities higher than $\sim 0.1$ ($H$ band) are necessary to allow the fringe-tracker FINITO\footnote{acronym of \textit{Fringe-tracking Instrument of NIce and TOrino}} to work under optimal conditions \citep{gai04}. 
These two contradictory constraints impose the usable range of the spatial frequencies ${f=B/ \lambda}$, where $B$ is the baseline length and $\lambda$ the observation  wavelength, to be that associated with the second lobe of the Uniform-Disc (UD) visibility function. 
In the following, we use the term \textit{resolution criterion} to summarise these constraints. They require that the maximum value of the dimensionless parameter ${z=\upi \phi f}$, where $\phi$ is the angular diameter, remains between 3.832, where the first zero of the uniform-disc visibility function appears, and 7.016 (second zero). For instance, observations in the $K$ band with AMBER of scientific targets with angular diameters of \unit[10]{mas} impose to the longest VLTI baseline length to be between 55 and \unit[101]{m} (first and second zero). The choice of the $K$ band is driven by the presence of the strong CO first overtone transition around \unit[2.33]{$\umu$m}, allowing to probe different layers in the photosphere within the same filter. 

To increase the confidence in the  measurements, we record multiple observations of each target per observing night. This observing procedure ensures obtaining sufficient amount of data to compensate for fringe-tracking deficiency, occurring when contrast is low or under poor-seeing conditions. 

According to our resolution criterion, we select the scientific targets from the two catalogues CHARM2 \citep{richichi05a}, and CADARS \citep{pasinetti01}), which compile angular-diameter values derived from various methods. In order to observe them with similar instrumental configurations, we choose stars with approximately the same angular diameter ($\unit[\sim 10]{mas}$). 
The suitable calibrators are given in Table~\ref{TabObsSci}. Since the angular diameter of some of them is not found in the calibrator catalogues, we had to derive it from the fit of \textsc{marcs} + \textsc{turbospectrum} synthetic spectra on spectrophotometric measurements \citep{cruzalebes10,cruzalebes13}. For reasons of homogeneity, we applied this procedure to all calibrators.

%
\setcounter{table}{1}
\begin{table}
\caption{Fundamental parameters used as entries in the \textsc{marcs} models.}
\label{TabMarcsSci} 
\centering
   \begin{tabular}{lcccccc}
  \hline 
 Target(s) & $T_\mathrm{eff}$~(K) & $\log g$ & $\mathcal{M}/\mathcal{M}_{\odot}$  & [Fe/H] & [$\alpha$/Fe] & C/O \\ 
  \hline
   $\alpha$~Car & 7000 & 2.0 & 5.0 & 0.0 & 0.0 & 0.54 \\
   $\beta$~Cet & 4660 & 2.1 & 1.0 & 0.0 & 0.0 & 0.54 \\
   $\alpha$~TrA & 4350 &  1.15 & 2.8 & 0.0 & 0.0 & 0.54 \\
   $\alpha$~Hya & 4300 & 1.3 & 1.1 & 0.0 & 0.0 & 0.54 \\
   $\zeta$~Ara & 4250 & 1.9 & 1.8 & -0.5 & 0.2 & 0.54 \\
   $\delta$~Oph & 3650 &  1.3 & 1.2 & 0.25 & 0.0 & 0.54 \\
   $\gamma$~Hyi & 3500 & 1.0 & 1.0 & 0.0 & 0.0 & 0.54 \\
   $o_{1}$~Ori & 3450 & 0.8 & 0.9 & 0.0 & 0.0 & 0.54 \\
   $\sigma$~Lib & 3450 & 0.8 & 0.9 & 0.0 & 0.0 & 0.54 \\
   $\gamma$~Ret & 3450 & 0.8 & 0.9 & 0.0 & 0.0 & 0.54 \\
   CE~Tau & 3400 & 0.0 & 12.0 & 0.0 & 0.0 & 0.54 \\
   T~Cet & 3250 & -0.5 & 7.0 & 0.0 & 0.0 & 0.54 \\
   TX~Psc & 3000 & 0.0 & 2.0 & -0.5 & 0.2 & 1.02 \\
   R~Scl & 2600 & 0.0 & 2.0 & 0.0 & 0.0 & 1.35 \\
   W~Ori & 2600 & 0.0 & 2.0 & 0.0 & 0.0 & 1.17 \\
   TW~Oph & 2600 & 0.0 & 2.0 & 0.0 & 0.0 & 1.17 \\
   \hline
  \end{tabular}
\end{table}

\subsection{Observation logbook} \label{SubSecLog}

A sample of 16 cool stars:~10~O-rich giants, 2~supergiants, and 4~C-rich giants, were observed in May~2009 (3~nights), August~2009 (2~nights), November~2009 (3~nights), March~2010 (3~nights), and December~2010 (4~nights), using the AMBER instrument at the focus of the ESO/VLTI, with three auxiliary telescopes (ATs). 
All observations were done using the Medium-Resolution-$K$-band (MR-K) spectral configuration, centered on ${\lambda=\unit[2.3]{\umu\mathrm{m}}}$, providing about 500 spectral channels with ${\mathscr{R}=1500}$. The observation logbook is given in Cruzal\`ebes et al. (submitted to MNRAS).
  
\section{Deriving the angular diameters} \label{SecAngdiam}

The true (calibrated) observables, defined hereafter, are derived from the AMBER output measurements, using the \textsc{spidast} modular software suite we have developed since 2006 \citep{cruzalebes08,cruzalebes10,cruzalebes13}. Recently made available to the community\footnote{\url{https://forge.oca.eu/trac/spidast}}, \textsc{spidast} performs the following automatised operations:~weighting of non-aberrant visibility and triple product data, fine spectral calibration at sub-pixel level, accurate and robust determinations of stellar diameters for calibrator sources, and of their uncertainties as well, correction for the degradations of the interferometer response in visibility and triple product, fit of parametric chromatic models on SPI observables, extraction of model parameters. 

We measure the angular diameter for each scientific target, by fitting synthetic limb-darkened brightness profiles on the visibility and the triple product. In an attempt to reproduce the behavior of the true observables, especially in the second lobe of the visibility function, we use the numerical Center-to-Limb Variation (CLV) profile w.r.t. the impact parameter, given by the \textsc{marcs} \citep{gustafsson08} + \textsc{turbospectrum} codes \citep{alvarez98,plez12}. 
       
%
\begin{figure}
\centering
   \includegraphics[scale=0.5, clip=true, trim=-30 0 0 0]{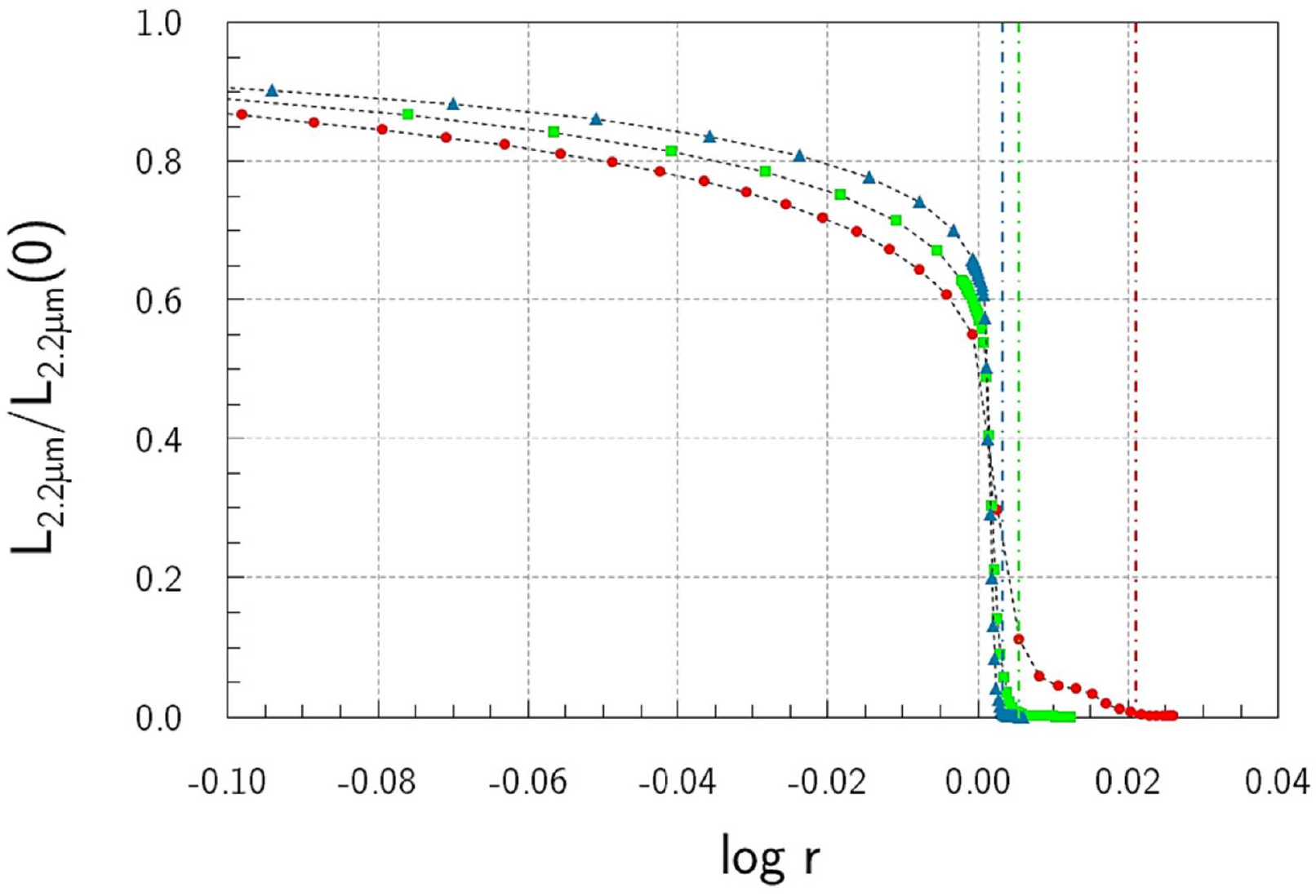}
\caption{Model CLV profiles at ${\lambda=\unit[2.2]{\umu\mathrm{m}}}$ for:~$\beta$~Cet (blue triangles); $\delta$~Oph (green squares); and TX~Psc (red circles). The dash-dot vertical lines show the values of the impact parameter at 0.5\% of the intensity:~1.008 for $\beta$~Cet, 1.013 for $\delta$~Oph, and 1.050 for TX~Psc ($r$ is in Rosseland radius unit).}
\label{Fig:intens22}  
\end{figure} 

\subsection{Computing reliable uncertainties} \label{SubSecUncert}

For each Observing Block (OB), the angular diameter is given by the modified gradient-expansion algorithm \citep{bevington92}, a robust fitting technique based on the minimisation of the weighted $\chi^{2}$, and adapted from \citet{marquardt63}. As ``robust'', we mean a final result insensitive to small departures from the model assumptions from which the estimator is optimised \citep{huber09}. We improve the robustness of the results of the fitting process by removing input measurements with low SNR ($<3$), as well as values considered as extremal residuals, i.e. showing exceedingly large discrepancies with the model.

%
   \begin{figure*}
   \centering
   \includegraphics[scale=1.1, clip=true, trim=0 0 0 -10]{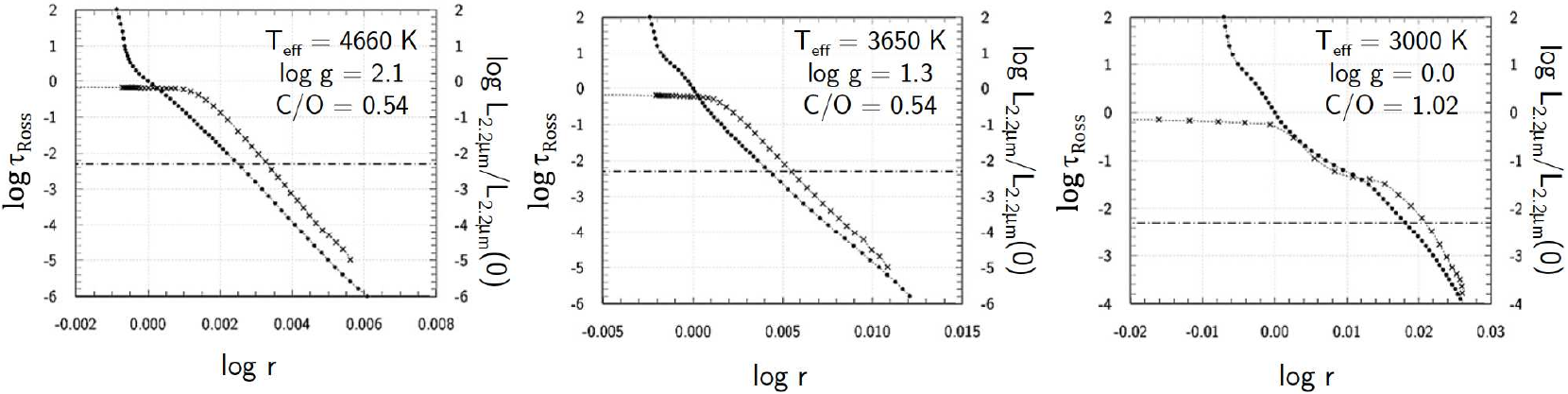}
   \caption{Rosseland optical depth (dots, left axis) and CLV profile at ${\lambda=\unit[2.2]{\umu\mathrm{m}}}$ (crosses, right axis) w.r.t. impact parameter (in Rosseland radius unit) for different \textsc{marcs} models (left and central panels:~O-rich stars; right panel:~C-rich). Dash-dot horizontal lines:~limit of instrumental sensitivity (0.5\%).}%
              \label{FigOptDepth}%
    \end{figure*}

Since the data used for the fit are obtained from a complex cross-calibration process, we cannot ensure that the final uncertainties follow a Normal distribution, but the $\chi^{2}$ function remains usable as merit function for finding the best-fit model parameters. However, the formal output-parameter uncertainties, deduced from the diagonal terms of the covariance matrix of the best-fit parameters, give irrelevant and usually underestimated values \citep{press07,enders10}.
In our study, we deduce reliable uncertainties from the boundaries of the 68\% confidence interval of the residual-bootstrap distribution of the best-fit angular diameters \citep{efron79,efron82,cruzalebes10}. 

\subsection{Choosing the model input parameters} \label{SubSecInput}

Table~\ref{TabMarcsSci} lists the stellar parameters of the science targets:~effective temperature, surface gravity, and mass, of the \textsc{marcs} models used in the regression process, with the microturbulence parameter $\xi_\mathrm{turb}$=\unit[2]{km~s$^{-1}$}. We derive these parameters from the two-dimensional B-spline interpolation of the tables of $\log T_\mathrm{eff}$ \citep{dejager87}, $\log g$ \citep{allen01}, and $\mathcal{L} / \mathcal{L}_{\odot}$ \citep{allen01},  w.r.t. the spectral type. 

Because this method relies on the spectral type, which carries some level of subjectivity, we concede that it is probably not the 
most accurate method for the determination of fundamental stellar parameters (see also the
discussion in relation with Fig.~\ref{FigTeff} in Sect.~\ref{SecHR}).
However, this method, currently used to measure the angular diameters of interferometric calibrators \citep{borde02,cruzalebes10}, provides a homogeneous way to convert various spectral types into fundamental parameters, all over the HRD.
In Sect.~\ref{SecSensi}, we investigate the sensitivity of the angular diameters to the adopted model stellar parameters, and show that this sensitivity is not an issue.

\subsection{Fitting the model limb-darkened intensity} \label{SubSecMarcsmod}

The spherically symmetric \textsc{marcs} model atmospheres, assuming local thermodynamic and hydrostatic equilibrium, are characterised by the following parameters:~effective temperature $T_\mathrm{eff}$,  surface gravity $g$, and mass $\mathcal{M}$, with ${g=G\mathcal{M}/\mathcal{R}_\mathrm{Ross}^2}$, where $\mathcal{R}_\mathrm{Ross}$ is the radius at ${\tau_{\mathrm{Ross}}=1}$. Using the \textsc{turbospectrum} code\footnote{in the $K$-band, we include all isotopomers of CO, C2, CN, as well as H$_{2}$O$^{16}$, and the atomic lists extracted from Uppsala-VALD}, we compute CLVs of the monochromatic radial intensity $L_{\lambda}(r)$ (also called \textit{spectral radiance}, in W~m$^{-2}$~$\umu$m$^{-1}$~sr$^{-1}$), where ${r=\mathcal{R}/\mathcal{R}_\mathrm{Ross}}$ is the impact parameter. Figure~\ref{Fig:intens22} shows the CLV profiles, at ${\lambda=\unit[2.2]{\umu\mathrm{m}}}$, calculated with \textsc{turbospectrum} using a \textsc{marcs} model, with three different sets of input parameters (Table~\ref{TabMarcsSci}), associated to~:~$\beta$~Cet (O-rich star, blue triangles); $\delta$~Oph (O-rich star, green squares); and TX~Psc (C-rich star, red circles). 

According to the Van Cittert-Zernike theorem \citep{goodman85}, the monochromatic synthetic visibility of a centro-symmetric brightness distribution of angular diameter $\phi$ is 
\begin{equation}
   {V}_\lambda \left( \phi \right) = 2\upi \frac{\left| \int^{r_\mathrm{out}}_{0} L_{\lambda}\left( r \right) \: 
   \mathrm{J}_{0}\left( \upi r \phi \frac{B}{\lambda} \right) \: r dr \right|}{M_{\lambda}}  ,
   \label{ModelVisLD} 
\end{equation}
where J$_{0}$ is the Bessel function of the first kind of order 0, $r_\mathrm{out}$ is the dimensionless parameter defined as ${r_\mathrm{out}=\mathcal{R}_\mathrm{out}/\mathcal{R}_\mathrm{Ross}}$, where $\mathcal{R}_\mathrm{out}$ is the outer radius, and ${M_{\lambda}=2\upi~{\int^{r_\mathrm{out}}_{0} L_{\lambda}\left( r \right) \: r dr}}$ is the monochromatic flux (also called \textit{spectral radiant exitance}, in W~m$^{-2}$~$\umu$m$^{-1}$). Numerical integration from 0 to $r_\mathrm{out}$ is performed using the trapezoidal rule on a grid, with a step width decreasing from the centre to the limb. 

%
   \begin{figure*}
   \centering
   \includegraphics[scale=0.8,clip=true, trim=0 0 0 0]{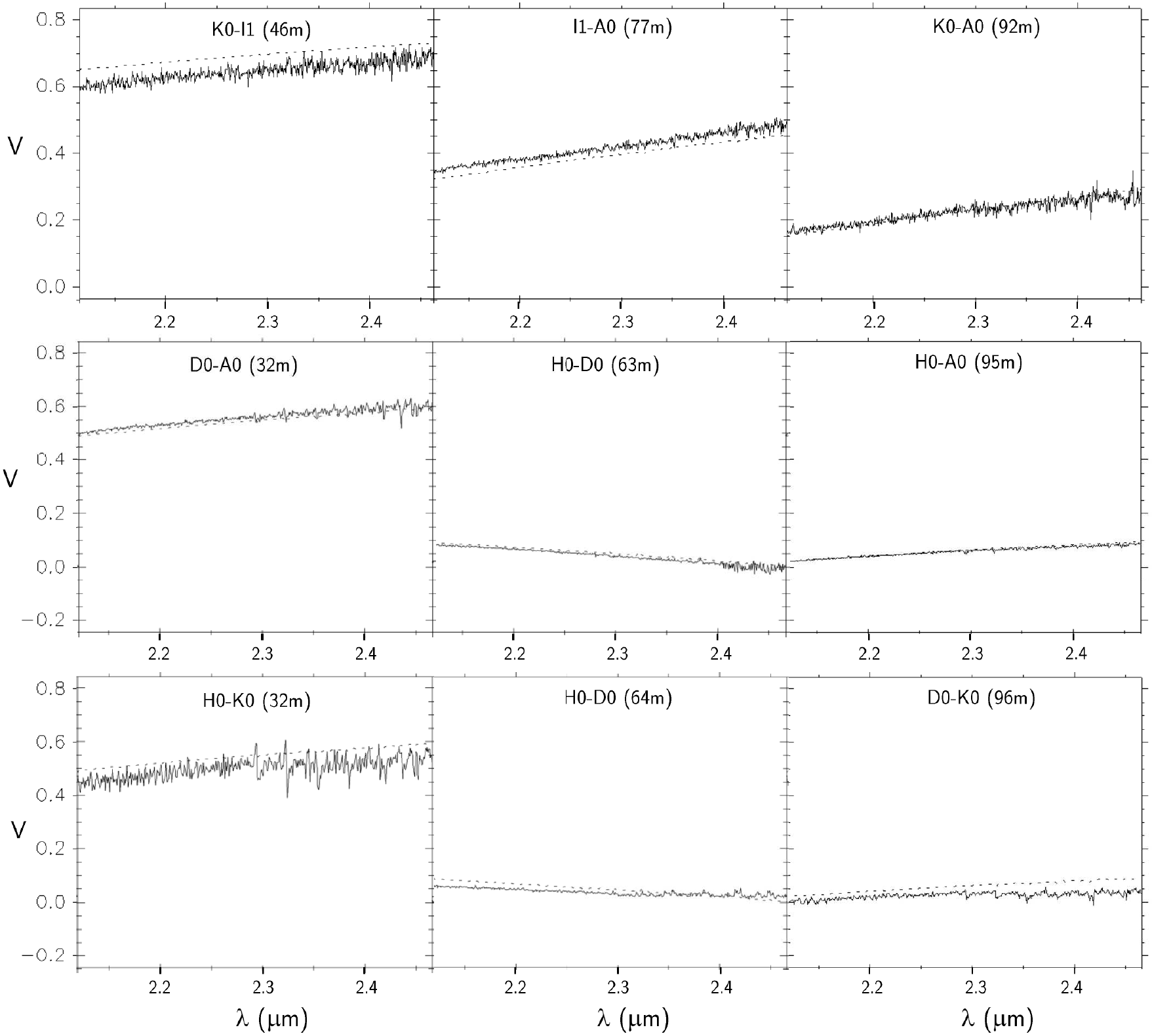}
    \caption{Three examples of \textsc{marcs}-CLV fitting results, obtained with visibility measurements:~$\beta$~Cet in top panels (MJD=55\,443.16); $\delta$~Oph in middle panels (MJD=54\,975.24); TX~Psc in bottom panels (MJD=55\,143.07). Baselines are projected on the sky. The model visibility profiles at medium spectral resolution (${\mathscr{R}=1500}$) are in short dashes. The measured median absolute uncertainties in visibility are (from left to right):~for $\beta$~Cet:~0.02, 0.03, and 0.02; for $\delta$~Oph:~0.010, 0.004, and 0.003; and for TX~Psc:~0.015, 0.003, and 0.004.}
             \label{AllFit}%
    \end{figure*}

%
\begin{table}
\caption{Sensitivity of the angular diameter $\phi$ (in mas) w.r.t. the model parameters (gravities are in c.g.s.).}
\label{TabSensib} 
\centering
\begin{tabular}{lccc}
  \hline 
& $T_\mathrm{eff}$~(K) & $\log~g=1.05$ & $\log~g=1.9$ \\
  \hline
&  3950  & $\begin{cases} \phi=7.060(2) \\ \chi^{2}=2.03 \end{cases}$ &  $\begin{cases} \phi=7.083(2) \\ \chi^{2}=1.98 \end{cases}$ \\
$\zeta$~Ara \\
&  4250 & $\begin{cases} \phi=7.039(2) \\ \chi^{2}=2.01 \end{cases}$ &  $\begin{cases} \phi=7.066(1) \\ \chi^{2}=1.83  \end{cases}$ \\
  \hline
  \\ \\
  \hline 
& $T_\mathrm{eff}$~(K) & $\xi_\mathrm{turb}$=\unit[2]{km~s$^{-1}$} & $\xi_\mathrm{turb}$=\unit[5]{km~s$^{-1}$} \\
  \hline
 TX~Psc & 3000 & $\begin{cases} \phi=10.053(2) \\ \chi^{2}=21.5 \end{cases}$ &  $\begin{cases} \phi=9.986(2) \\ \chi^{2}=22.5 \end{cases}$ \\
  \hline
 \end{tabular}
\end{table}

To be accurately evaluated, the integral on $r$ in Eq.(\ref{ModelVisLD}) requires $L_{\lambda}(r)$  to be extended all the way to a value of $r_\mathrm{out}$ corresponding to the lower boundary of the sensitivity of the AMBER instrument. In the $K$ band, this boundary has been measured around 0.5\% of the maximum emission \citep{duvert10,absil10}. With the \textsc{marcs} models, the lower boundaries of the
Rosseland optical depth, used to compute the intensity distributions $L_{\lambda}(r)$, are ${\tau_{\mathrm{Ross}}=10^{-6}}$ for O-rich stars, and ${\tau_{\mathrm{Ross}}=10^{-4}}$  for C-rich stars (Fig.~\ref{FigOptDepth}). Thus, ${\mathcal{R}_\mathrm{out}=\mathcal{R}\left(\tau_{\mathrm{Ross}}=10^{-6}\right)}$ for O-rich stars, and ${\mathcal{R}_\mathrm{out}=\mathcal{R}\left(\tau_{\mathrm{Ross}}=10^{-4}\right)}$ for C-rich stars. These bottom levels ensure that the blanketing is correctly taken into account, and that the thermal structure in the line-forming region remains unchanged w.r.t. atmospheres that would be computed with even smaller optical-depth boundaries. According to Fig.~\ref{FigOptDepth}, these optical-depth lower boundaries are associated with intensity levels of $< 10^{-5}$ and $\sim 10^{-4}$, respectively, thus far below the instrumental sensitivity, as it should be to work in safe conditions.
Moreover, in Sect.~\ref{SecSensi} we evaluate the sensitivity of the angular diameter to the \textsc{marcs} model used to compute the CLV. 

Figure~\ref{AllFit} shows three typical results of the \textsc{marcs}-CLV fits obtained with the visibility measurements of individual OBs, for $\beta$~Cet, $\delta$~Oph, and TX~Psc. For the sake of clarity, the true visibility values are shown without error bars. 
In addition to the fit of the \textsc{marcs}-CLV profile on the true visibilities, we also compute the angular diameter using fits on triple product data (Table~\ref{TabLogFinal}). 

\section{Studying the sensitivity to model parameters} \label{SecSensi}

The effective temperature, surface gravity and stellar mass adopted for the \textsc{marcs} model representing a given star are derived from the spectral type (Sect.~\ref{SubSecInput}). Unfortunately, neither the gravity, nor the stellar mass are strongly constrained by the spectral type alone. Therefore, there is a disagreement between the \textsc{marcs}-model parameters and the true stellar values. In this section, we study, for the 2 targets:~$\zeta$~Ara (K-giant), and TX~Psc (carbon star), the sensitivity of the angular diameter, to a change of input parameter values, such as:~$T_\mathrm{eff}$, $\log g$, $\xi_\mathrm{turb}$. 
 
Table~\ref{TabSensib} shows the sensitivity to the model parameters of the angular diameter, deduced from the fit on visibility data. The top table is for $\zeta$~Ara observed at MJD=54\,975.36, and the bottom table is for TX~Psc observed at MJD=55\,143.10. The uncertainties in angular diameter are the formal 1-$\sigma$ fitting errors. The choice of the different values of $T_\mathrm{eff}$ and $\log g$ used for this analysis are based on the typical uncertainties, \unit[300]{K} and \unit[1]{dex} respectively, the latter coming from the \textit{a posteriori} determination of the gravity (Sect.~\ref{SecHR} and Fig.~\ref{Fig:comp_log_g}). The sensitivity to $\xi_\mathrm{turb}$ is studied with the values 2 and \unit[5]{km~s$^{-1}$}, for the carbon star.

The highest deviations from the nominal values of the angular diameter, i.e. \unit[0.03]{mas} for $\zeta$~Ara and \unit[0.07]{mas} for TX~Psc, are smaller than the final uncertainties, \unit[0.12]{mas} and \unit[0.36]{mas} respectively (Table~\ref{TabLogFinal}). 
Although we cannot infer quantitative general sensitivity rules from only two examples, our results show that such changes as \unit[300]{K} for $T_\mathrm{eff}$, roughly \unit[1]{dex} for $\log g$ and a factor of two for $\xi_\mathrm{turb}$ induce variations on the final angular diameter which are smaller than its absolute uncertainty.

\section{Studying the temporal variability of the angular diameter} \label{SecTempoVar}

To study the temporal variability of the angular diameter, we group together the observing blocks of the same observing epoch over consecutive days, for each scientific target. Table \ref{TabLogFinal} gives the best-fit angular diameters of the scientific targets, separately for each observation epoch and for the average over all runs. MJD is the Modified Julian Day for the middle of each observing period. The notations $\phi_\mathrm{V}$ and $\phi_\mathscr{T}$ stand for the weighted means of the angular diameters resulting from fits of the \textsc{marcs} CLVs on visibility and triple product data, respectively.

\begin{figure*}
   \centering
   \includegraphics[scale=0.95,clip=true, trim=0 0 0 0]{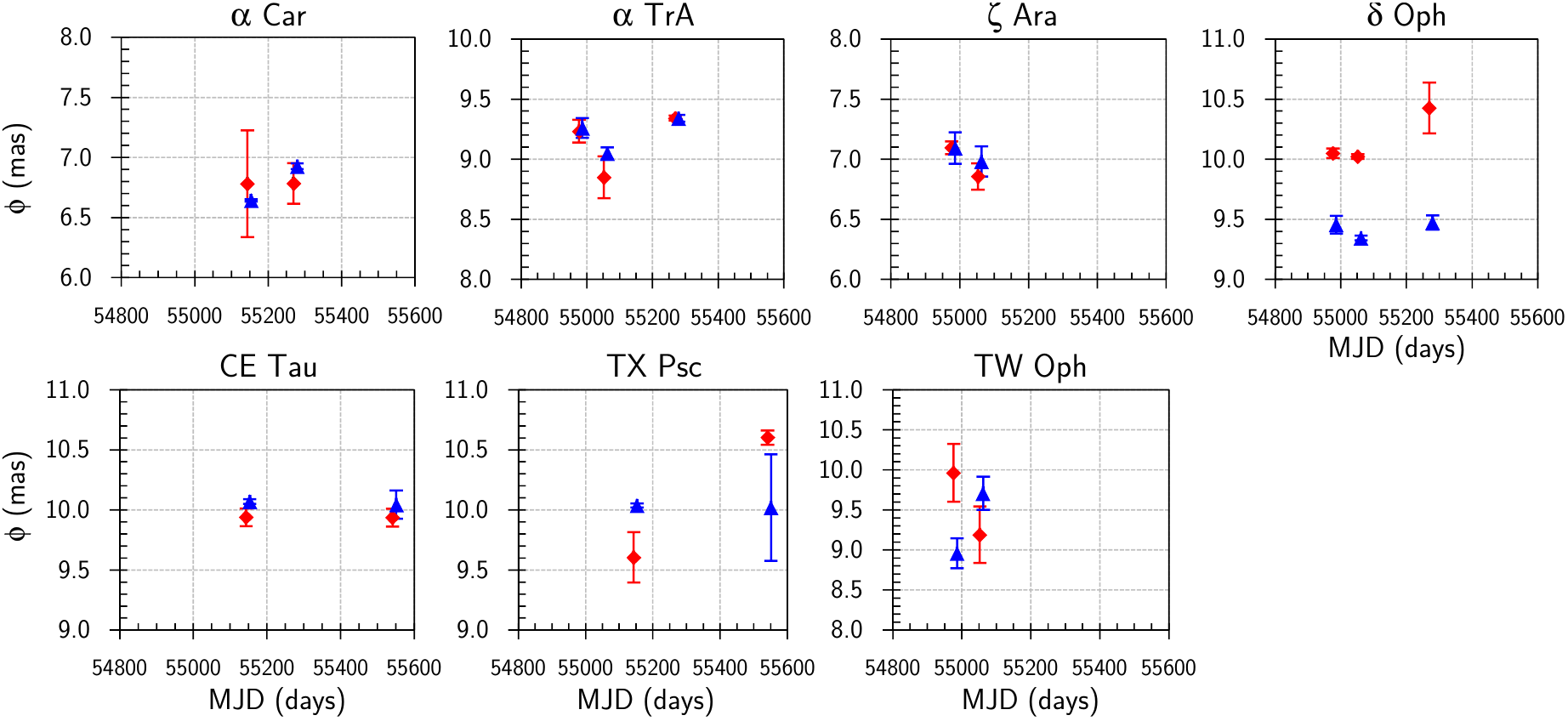}
    \caption{Temporal behaviour of the best-fit angular diameter, from visibilities (red diamonds) and triple products (blue triangles). Top panels:~science targets showing no photometric variation; bottom panels:~targets known as photometric variables (i.e., with a GCVS entry). The symbols which mark the results associated with the same observing epoch are slighly shifted horizontally, in order to separate the error bars. }
             \label{MJDphi}%
    \end{figure*}

Figure~\ref{MJDphi} shows the temporal behaviour of the best-fit angular diameter, for our scientific targets observed over different epochs. Except for TX~Psc, which shows two different values of $\phi_\mathrm{V}$, but not of $\phi_\mathscr{T}$, we find no evidence for temporal variation of the angular diameter for our targets, given the uncertainties. 
To perform a meaningful study of the angular-diameter time variability, a 
larger amount of data would have been needed for our targets. Unfortunately, we did not 
succeed in convincing the Observing Programmes Committee to allow 
supplementary observing time for this purpose.

\begin{table}
  \caption{Best-fit angular diameters derived from the visibility and the triple product, for each observation epoch, and final angular diameters of the science targets, after averaging over all OBs. MJD is the Modified Julian Day for the middle of each observation period.}
  \label{TabLogFinal}
\centering
  \begin{tabular}{llccc} 
  \hline 
 Name & MJD~(days) & $\phi_\mathrm{V}$~(mas) & $\phi_\mathscr{T}$~(mas) & $\phi_\mathrm{final}$~(mas)\\ 
   \hline
   \hline
$\alpha$~Car & 54\,143.26 & 6.78(45) & 6.64(1) & 6.92(11) \\
             & 55\,269.14 & 6.78(17) & 6.93(2) & \\
   \hline
$\beta$~Cet  & 55\,541.17 & 5.84(40) & 5.45(5) & 5.51(25)\\
   \hline
$\alpha$~TrA & 54\,976.23 & 9.23(10) & 9.26(8) & 9.24(2)\\
             & 55\,052.12 & 8.85(18) & 9.05(5) & \\
             & 55\,269.32 & 9.34(2) & 9.34(3) & \\
   \hline
$\alpha$~Hya & 55\,269.23 & 9.37(5) & 9.35(7) & 9.36(6) \\
   \hline
$\zeta$~Ara  & 54\,976.25 & 7.10(5) & 7.09(13) & 7.09(12)\\
             & 55\,053.18 & 6.86(11) & 6.98(13) &  \\
   \hline
$\delta$~Oph & 54\,976.22 & 10.05(4) & 9.46(7) & 9.93(9)\\
             & 55\,051.99 & 10.02(2) & 9.34(2) & \\
             & 55\,269.88 & 10.43(21) & 9.47(6) & \\
   \hline
$\gamma$~Hyi & 55\,539.80 & 8.77(6) & 8.82(12) & 8.79(9)\\
   \hline
$o_{1}$~Ori  & 55\,144.24 & 8.93(15) & 10.04(5) & 9.78(10)\\
   \hline
$\sigma$~Lib & 55\,268.81 & 11.73(14) & 11.19(3) & 11.33(10) \\
   \hline
$\gamma$~Ret & 55\,539.83 & 7.44(2) & 7.44(2) & 7.44(2)\\
   \hline
CE~Tau       & 55\,143.28 & 9.94(7) & 10.07(2) & 9.97(8)\\
             & 55\,541.24 & 9.94(7) & 10.04(12) & \\
   \hline
T~Cet        & 55\,143.13 & 9.60(11) & 9.70(1) & 9.70(8)\\ 
   \hline
TX~Psc       & 55\,143.08 & 9.61(21) & 10.04(2) & 10.23(36)\\
             & 55\,541.07 & 10.60(6) & 10.02(45) & \\ 
   \hline
W~Ori        & 55\,143.72 & 9.62(1) & 9.79(7) & 9.63(4) \\
   \hline
R~Scl        & 55\,143.56 & 10.31(5) & 9.88(2) & 10.06(5)\\
   \hline
TW~Oph       & 54\,976.35 & 10.59(38) & 9.53(20) & 9.46(30)\\
             & 55\,052.19 & 9.19(35) & 9.71(21) & \\
\hline
\end{tabular}
\end{table}

\section{Computing the final angular diameter} \label{SecFinStellarDiam}

Rather than applying a global fit on all data sets \citep[see e.g.,][]{lebouquin08,domiciano08}, which is the commonly used method with VLTI/AMBER data, we propose to combine multiple measurements obtained for a given star under different instrumental and environmental circumstances \citep[see e.g.,][]{ridgway80,richichi92,dyck96a}. Using the visibility and the triple product, we compute the angular diameter averaged over all OBs, with a weighting factor derived from the uncertainty on the angular diameter, the quality of the fit, and the seeing conditions during each OB. Then, we combine the two angular-diameter values, which leads to a unique final value (last column of Table~\ref{TabLogFinal}). We note that $\delta$~Oph is the only star for which $\phi_\mathrm{V}$ and $\phi_\mathscr{T}$ are significantly different (up to 10\%, as seen in Table~\ref{TabLogFinal} and Fig.~\ref{MJDphi}), although we have no explanation for that discrepancy. 

\section{Confronting our results with those of the literature} \label{SecDiscuss}

Here, we compare our final angular-diameter values with those derived from measurements obtained by other instruments or methods. Table~\ref{TabPubDiamSci} gathers the values published in the literature, related to limb-darkened models, derived from indirect methods, Lunar Occultation (LO), and LBI. 
These values are obtained in various spectral ranges and related to various photospheric models. Their large dispersions make them difficult to use in a direct comparison with our results, which are repeated in the $\phi_\mathrm{LBI}$ column under Ref.~(48). 
Therefore, we believe that the only meaningful comparison is between our 
values and those from the literature obtained with LBI in the same spectral domain ($K$ band), 
as done in the last column of Table~\ref{TabPubDiamSci}.

Apart for TX~Psc, only small differences are found between our new values and the published LBI values for the seven science targets:~$\alpha$~Car, $\beta$~Cet, $\alpha$~Hya, $\delta$~Oph, CE~Tau, W~Ori, and R~Scl. Such a good agreement supports the validity and the reliability of our method, which gives, in addition, reliable uncertainties. Our study provides the first LBI determinations of the angular diameter for the eight other targets:~$\alpha$~TrA, $\zeta$~Ara, $\gamma$~Hyi, $o_{1}$~Ori, $\sigma$~Lib, $\gamma$~Ret, T~Cet, and TW~Oph.   

Coming back to TX~Psc, this star has often been observed in the past using high-resolution techniques, giving an angular diameter slightly larger than our new measurement. Given the error bars, our value is in good agreement with the value from \citet{barnes78}, derived from the visual surface brightness method. 
\citet{richichi95} attribute to the temporal
variability of $\phi$ already noted previously for TX~Psc (a Lb-type variable) 
most of the disagreement between
their LO measurement and the LBI values of \citet{quirrenbach94a}, obtained in the red part of the visible spectral domain with
the MkIII Optical Interferometer, and of \citet{dyck96b}, obtained
at \unit[2.2]{$\umu$m} with the IOTA interferometer.
From repeated measurements, Quirrenbach et al. suggested a substantial variation of the angular diameter, correlated with the visual magnitude, varying from 4.8 to 5.2 in \unit[220]{days} \citep{watson06}. As shown in Sect.~\ref{SecTempoVar}, our data tend to confirm this variation.

\section{Hertzsprung -- Russell Diagram} \label{SecHR}

In this section, we use the values of the angular diameters of our calibrators and science targets to infer their location in the HRD ($T_\mathrm{eff}$ -- $\mathcal{L}$).

The luminosity $\mathcal{L}$ is defined, in the \textsc{marcs} models, from the relation ${\mathcal{L}=4 \upi \mathcal{R}_\mathrm{Ross}^2 F\left(\mathcal{R}_\mathrm{Ross}\right)}$,
where $F\left(\mathcal{R}_\mathrm{Ross}\right)$ is the flux per unit surface emitted by the layer located at the Rosseland radius \citep{gustafsson08}. The effective temperature $T_\mathrm{eff}$ is then defined according to ${F\left(\mathcal{R}_\mathrm{Ross}\right)=\sigma T_\mathrm{eff}^4}$.

We convert the best-fit angular diameter $\phi$ into an empirical Rosseland radius $\mathcal{R}_\mathrm{obs}$ thanks to the parallax $\varpi$. For the calibrators, $\phi$ is given by the fit of the model spectrum on the flux data. For the science targets, $\phi$ is given by the fit of the CLV profile on the SPI data. Thus, we compute the empirical luminosity $\mathcal{L}_\mathrm{obs}$ using the logarithmic formula 
  \begin{equation}
  \log \frac{\mathcal{L}_\mathrm{obs}}{\mathcal{L_{\odot}}} \approx 4 \log T_{\mathrm{eff}} + 2 \log \frac{\phi}{\varpi} - 10.984(7),
      \label{logL}
  \end{equation} 
where $T_\mathrm{eff}$ is in Kelvin, using the solar values $T_{\mathrm{eff},\odot}$=\unit[5777(10)]{K}  \citep{smalley05}, and $\mathcal{R}_{\odot}$=\unit[0.004\,6492(2)]{AU} \citep{brown98,amsler08}.

Table~\ref{TabFinal} gives the final fundamental parameters of our science
targets and calibrators. The uncertainty-propagation formulae given by \citet{winzer00}, based on the second-order Taylor approximation, are used
to compute the uncertainties on the derived fundamental parameters. 
For input uncertainties larger than 30\%, we use
the confidence interval transformation principle \citep[see e.g.,][]{smithson02,kelley07}. 

To ensure consistency with the fitting process,
which uses as model input parameters those derived from
the spectral type (\citealt{cruzalebes13} and Table~\ref{TabMarcsSci}), the value
adopted for the effective temperature of the star is the value listed
in Table~\ref{TabMarcsSci}. 
  
\begin{figure}
   \centering
   \includegraphics[scale=0.5, clip=true, trim=0 0 0 0]{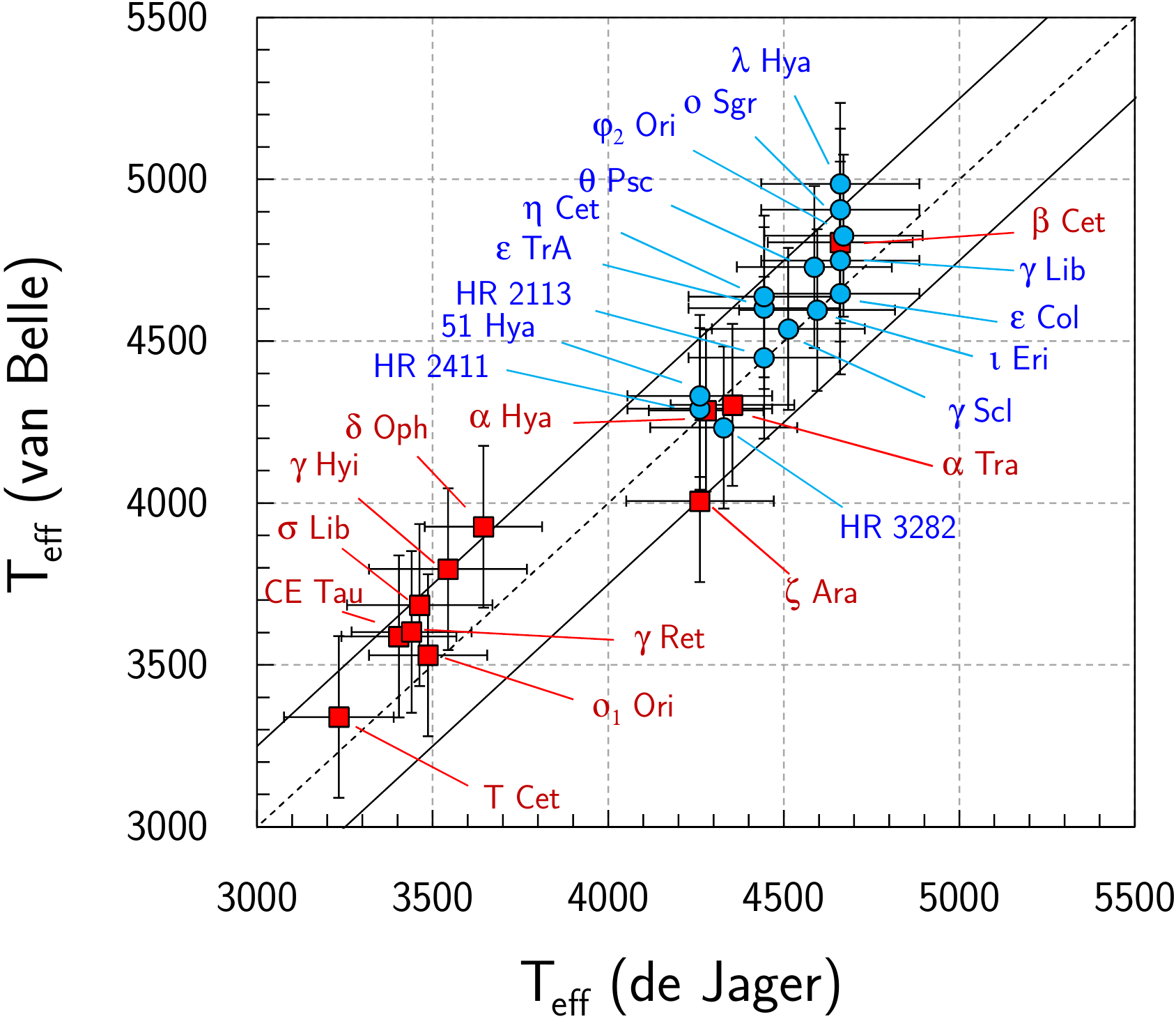}
   \caption{Effective temperatures deduced from the de-reddened $V-K$ color index (van Belle), versus from the spectral type (de Jager), for the targets of our observing sample. Red squares:~science targets. Blue dots:~calibrators. Solid lines:~\unit[$\pm$250]{K} thresholds.}%
              \label{FigTeff}%
    \end{figure}

To assess the accuracy of the value, we compare the effective temperature deduced from the spectral type for the giants and supergiants of types K and M, included in our samples of science and calibrator targets, with the temperature derived from the de-reddened $V-K$ index, using the empirical relationship provided by \citet{vbelle99b} 
     \begin{equation}
 T_\mathrm{eff}~\mathrm{(K)} \ = \ 3030 + 4750 \times  10^{-0.187\left(V-K\right)},
    \label{TeffVanBelle} 
   \end{equation}
where $2 < V-K < 9$. We find that the agreement between the effective temperatures, shown in Fig.~\ref{FigTeff}, is quite satisfactory, since their discrepancy is less than \unit[$\pm$300]{K}, which is of the same order than the absolute uncertainty given by the van Belle's formula (\unit[$\pm$250]{K}).
For the three carbon stars W~Ori, R~Scl, and TW~Oph, the adopted effective temperature of 
2600~K (Table~\ref{TabMarcsSci}) is consistent with the values derived by \citet{lambert86} with \unit[$\pm$100]{K} uncertainty:~respectively \unit[2680]{K}, \unit[2550]{K}, and \unit[2450]{K}.

\begin{figure}
   \centering
   \includegraphics[scale=0.45, clip=true, trim=0 0 0 0]{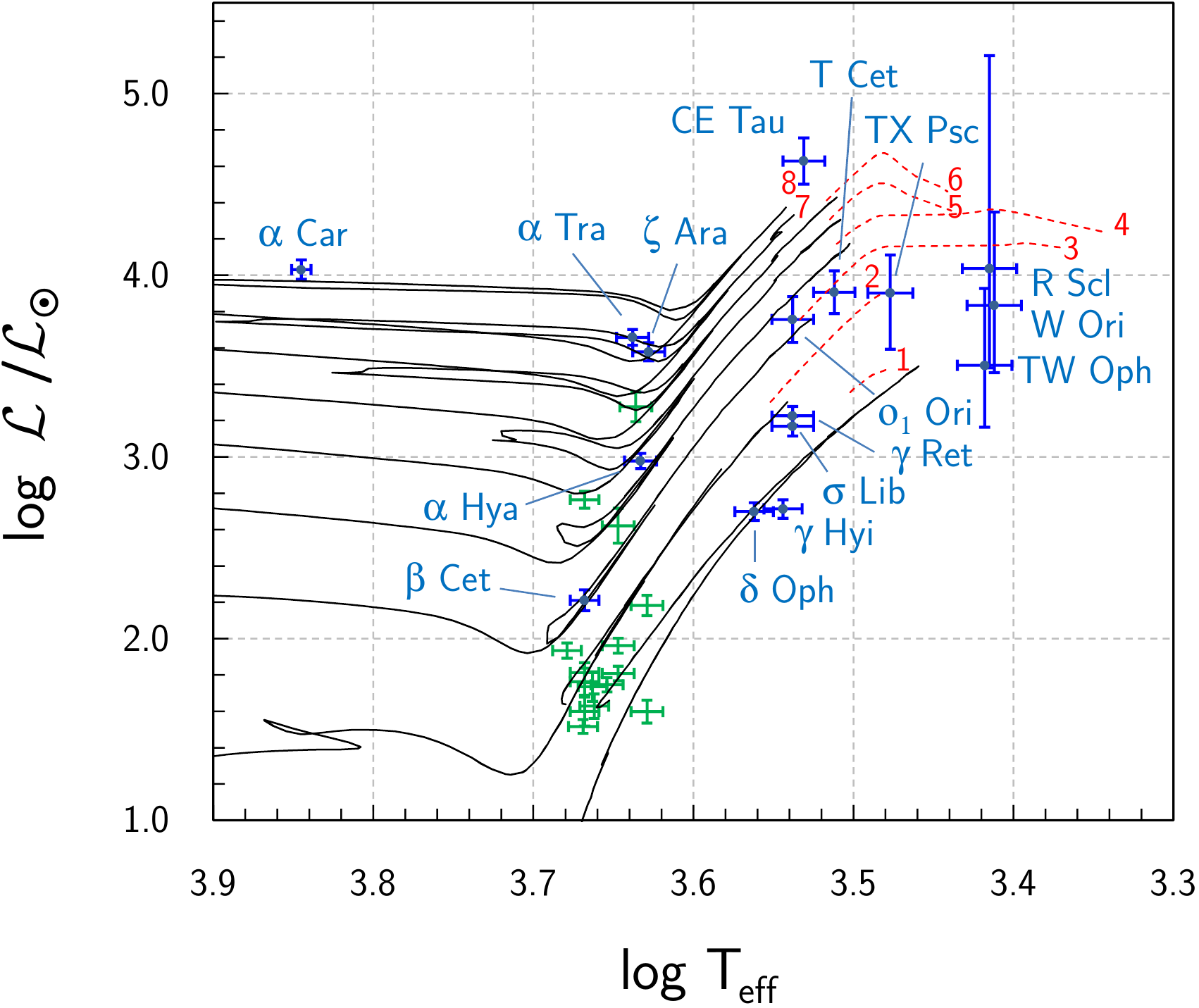}
      \caption{$T_\mathrm{eff}$ -- $\mathcal{L}_\mathrm{obs}$ diagram of the calibrators (thin green error bars) and science targets (thick blue error bars), with evolutionary tracks (black full lines) and asymptotic giant branches (red dashed lines), for different masses, indicated in red at the end of each track.}
         \label{FigTeffL}         
   \end{figure}

Figure~\ref{FigTeffL} shows the resulting $T_\mathrm{eff}$~--~$\mathcal{L}_\mathrm{obs}$ diagram, including the calibrators and the science targets. In order to distinguish between the error bars, the data points for R~Scl, W~Ori and TW~Oph are slightly shifted horizontally, although these 3 carbon stars have the same effective temperature \unit[2600]{K}.
This HRD displays as well evolutionary tracks from the Padova set \citep{bertelli08,bertelli09}, for ${Y=0.26}$ and ${Z=0.017}$, and for masses between 1 and 8~$\mathcal{M}_{\odot}$, where $Y$ is the helium abundance, and $Z$ the metallicity.

\begin{figure}
   \centering
   \includegraphics[scale=0.45, clip=true, trim=0 0 0 0]{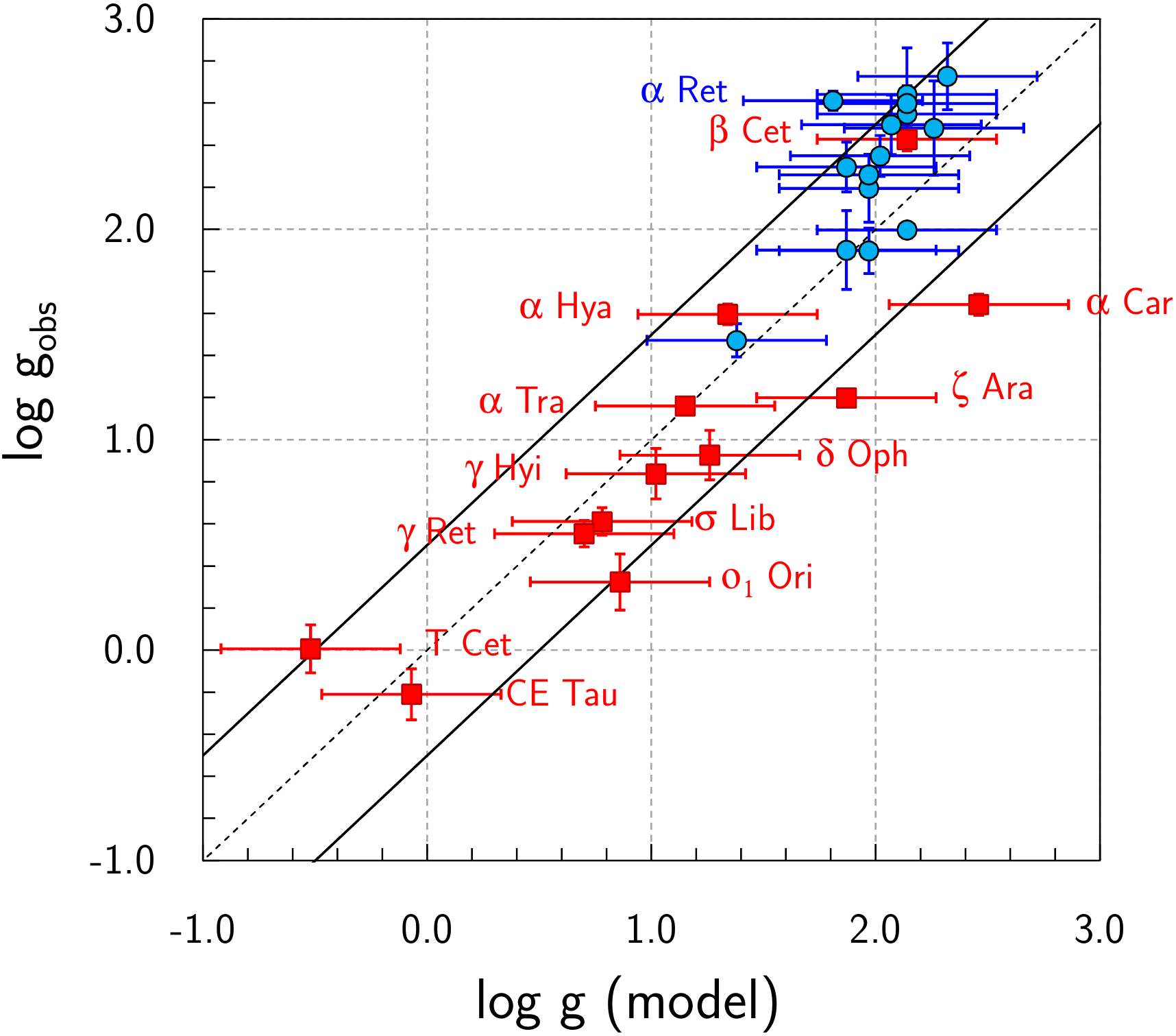}
      \caption{Comparison of the surface gravities used to select the \textsc{marcs} models, with the values $\log g_\mathrm{obs}$ derived from our final angular-diameter value. Red squares:~science targets; blue dots:~calibrators. The solid lines mark the \unit[$\pm 0.5$]{dex} thresholds.}   
         \label{Fig:comp_log_g}
   \end{figure}

These tracks make it possible to derive a rough estimate of the stellar mass $\mathcal{M}$, thus of the gravity $g_\mathrm{obs}$ at the Rosseland surface, deduced from the relation
\begin{equation}
   \log g_\mathrm{obs} \approx \log \frac{\mathcal{M}}{\mathcal{M}_{\odot}} - 2 \log \frac{\mathcal{R}_\mathrm{obs}}{\mathcal{R}_{\odot}} + 4.4374(5),
 \label{logg}
\end{equation}
using the value of the solar surface gravity given by \citet{gray05}. These mass and gravity values are also included in Table~\ref{TabFinal}. The comparison of the surface gravities $\log~g$, deduced from the spectral type and used to select the \textsc{marcs} models, with those derived {\it a posteriori} from the HRD, is done in Figure~\ref{Fig:comp_log_g}. We see that they agree within \unit[$\pm 0.5$]{dex}, except for the calibrator $\alpha$~Ret (${\log g-\log g_\mathrm{obs}=-0.80}$), and for the science targets $\alpha$~Car (+0.82), and $\zeta$~Ara (+0.67). Since the determination of the mass from the position along the evolutionary tracks in the HRD is well-constrained\footnote{we note, however, that the mass inferred from the HRD tracks corresponds to the \textit{initial} mass. But substantial mass loss along the evolution may significantly reduce the current mass below 
its initial value, implying that $g_\mathrm{obs}$ as we derive it could be somewhat overestimated}, we attribute the discrepancy in surface gravity to the ill-defined value derived from the spectral type and used for the model, at least for these 3 targets.

With the linear radius derived from the interferometry, and the luminosity following the relationship ${\mathcal{L}=4\upi\mathcal{R}^2\sigma T_\mathrm{eff}^4}$, the location of our targets in the HRD allows us to perform interesting checks of stellar structure related to the presence or absence of technetium, and to the Period -- Luminosity relationship.

\section{Technetium} \label{SecTc}

Technetium is an s-process element with no stable isotope that was first identified in the spectra of some M and S stars by \citet{merrill52}. With a laboratory half-life of \unit[$2.13 \times 10^{5}$]{yr}, the technetium isotope $^{99}$Tc is the only one produced by the s-process in thermally-pulsating AGB (TP-AGB) stars \citep[see][]{goriely00}. Due to the existence of an isomeric state of the $^{99}$Tc nucleus, the high temperatures encountered during thermal pulses strongly shorten the effective half-life of $^{99}$Tc  ($t_{1/2} \sim$ \unit[1]{yr} at \unit[$\sim 3 \times 10^{8}$]{K}) \citep{cosner84}, but the large neutron densities delivered by the $^{22}$Ne($\alpha$,n)$^{25}$Mg neutron source, operating at these high temperatures, more than compensate the reduction of the $^{99}$Tc lifetime \citep{mathews86}, and enable a substantial technetium production. 
The dredge-up episodes then carry technetium to the envelope, where it decays steadily
at its terrestrial rate of $t_{1/2}$=\unit[$2.13 \times 10^{5}$]{yr}. Starting from an
abundance associated with the maximum observed in Tc-rich AGB stars, technetium
should remain detectable during \unit[1.0 to $1.5 \times 10^{6}$]{yr} \citep{smith88}. If the dredge-up of heavy elements occurs after each thermal pulse, occurring every \unit[0.1 to $0.3 \times 10^{6}$]{yr}, virtually all s-process enriched TP-AGB stars should exhibit technetium lines.

This conclusion applies to the situation where the s-process is powered by the $^{22}$Ne($\alpha$,n)$^{25}$Mg neutron source operating in the thermal pulse itself. However, \citet{straniero95} advocated that the s-process 
nucleosynthesis mainly occurs during the interpulse with neutrons from 
$^{13}$C($\alpha$,n)$^{16}$O \citep[see][for a recent review]{kappeler11}.
When this process occurs in low-mass stars, and technetium is engulfed in the subsequent thermal pulse, it should not decay at a fast rate, because the arguments put forward by \citet{cosner84} and \citet{mathews86}, and discussed above, only apply to intermediate-mass stars with hot thermal pulses.

One thus reaches the conclusion that \textit{s-process-enriched} TP-AGB stars, of both low and intermediate mass, should necessarily exhibit technetium, unless 
the time span between successive dredge-ups become comparable to the Tc lifetime in the envelope. Indeed, all the S stars identified as TP-AGB stars by \citet{vaneck98} thanks to the Hipparcos parallaxes 
turned out to be Tc-rich, and a survey of technetium in the large Henize sample of S stars did not challenge that conclusion either \citep{vaneck99,vaneck00}. 

\begin{figure}
   \centering
\includegraphics[scale=0.45, clip=true]{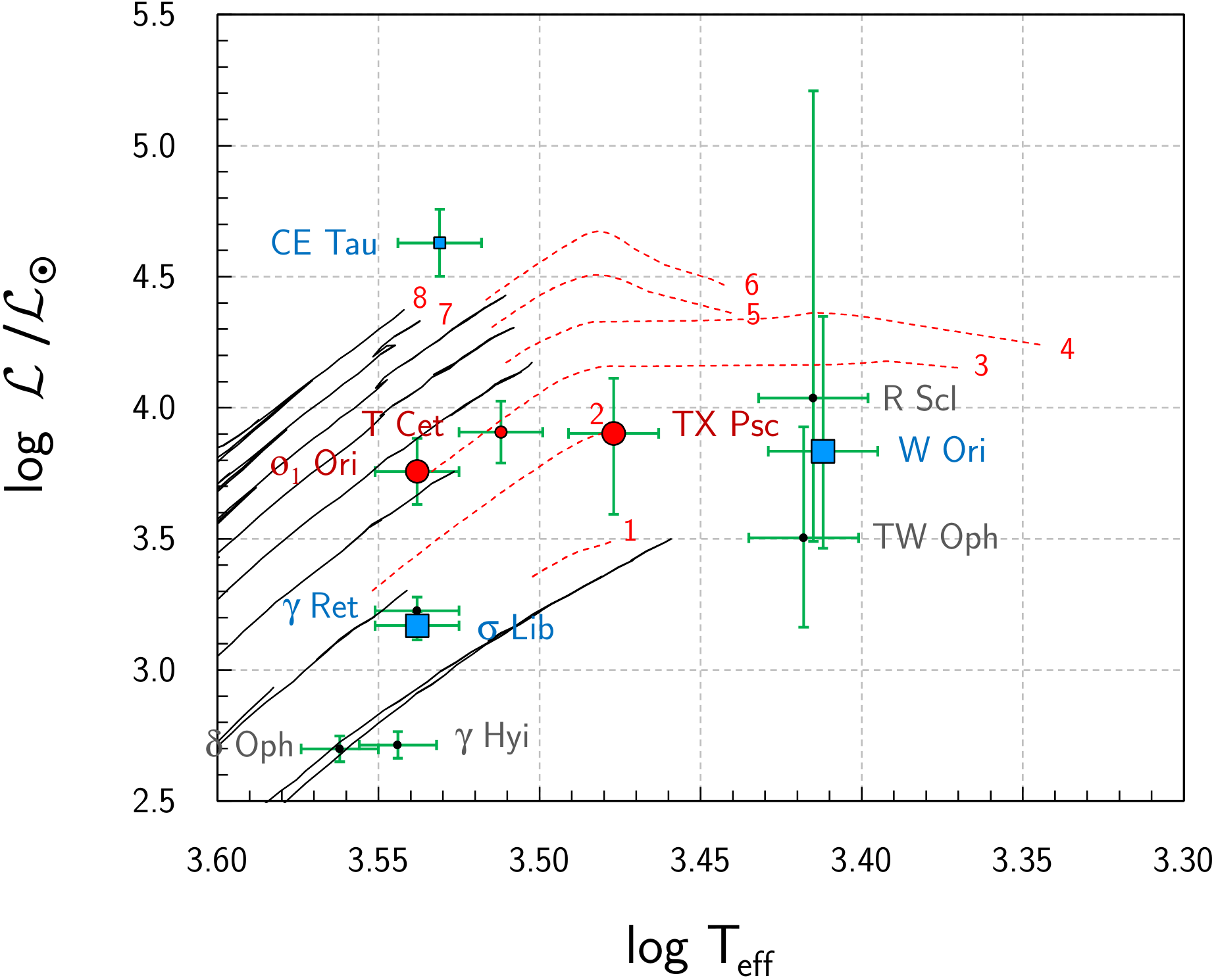}
      \caption{Same as Fig.~\ref{FigTeffL} restricted to the science targets, with their technetium content indicated. Large red circles:~Tc present; small red circle:~Tc probably present; large blue squares:~Tc absent; small blue square:~Tc doubtful; small black dots:~unknown Tc content. Red dashed lines: TP-AGB tracks, for masses indicated at the end of each curve.}
        \label{Fig:Tc}
   \end{figure}

The present sample allows us to check whether a similar conclusion holds true for a sample comprising oxygen-rich giants and supergiants, as well as carbon stars. 
The technetium content of our science targets has been collected from the literature (last column of Table~\ref{TabObsSci}), and displayed in graphical form in Fig.~\ref{Fig:Tc} where it is confronted to the TP-AGB tracks (dashed lines) for different stellar masses. The presence or absence of Tc conforms to the expectations that namely TP-AGB stars exhibit Tc, except for the carbon star W~Ori, where Tc has been tagged  as absent by two independent studies, despite the fact that this star lies well within the TP-AGB region, as it should for a cool carbon star anyway. The s-process content of that star has been studied by
\citet{abia02} who find only moderate s-process enhancements, if any  ($\le 0.3$~dex), and this fact alone may explain the absence of detectable Tc.
With the stellar parameters now available from our interferometric study for two more  carbon stars (R~Scl and TW~Oph) falling in that region 
of the HRD, it will be of interest to perform a similar analysis on these two stars to get constraints on their nucleosynthesis processes. 

\section{Period -- Luminosity relation} \label{SecPL}

\begin{figure}
   \centering
\includegraphics[scale=0.45, clip=true, trim=0 0 0 0]{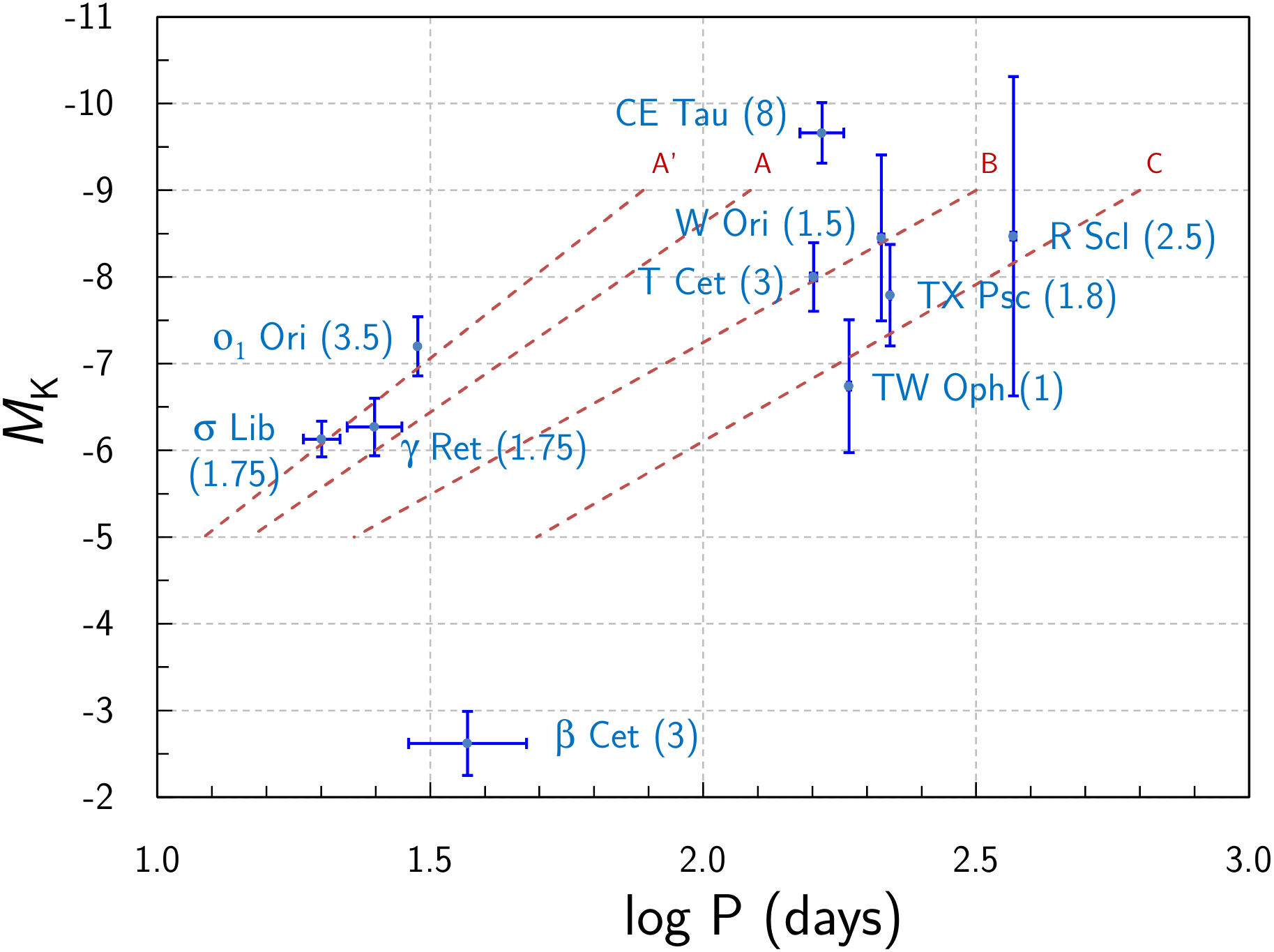}
      \caption{Period -- Luminosity diagram for science targets with available pulsation periods. $M_\mathrm{K}$ values are derived from the dereddened K magnitudes and the distance moduli. Stellar masses between parentheses  (in $\mathcal{M}_{\odot}$). Location of A' to C sequences taken from \citet{tabur10}.}
        \label{Fig:P-L}
   \end{figure}

Since the present study derives radii and masses for some semi-regular variables, we also derive the pulsation constant $Q$ \citep[e.g.,][]{fox82}, defined as
\begin{equation}
Q = P\; \left(\frac{\mathcal{M}}{\mathcal{M}_{\odot}}\right)^{1/2}\; \left(\frac{\mathcal{R}_\mathrm{obs}}{\mathcal{R}_{\odot}}\right)^{-3/2},
\label{PulsCst}
\end{equation}
where $P$ is the pulsation period given in Table~\ref{TabObsSci}. For pulsating stars with available periods of variation, we include $Q$ in Table~\ref{TabFinal} (last column).
Values of $Q$ smaller than \unit[0.04]{d} are typical of overtone pulsators \citep{fox82}. Indeed, in the Period -- Luminosity diagram ($M_\mathrm{K}, P$) shown in Fig.~\ref{Fig:P-L}, and following the terminology introduced by \citet{wood00}, most of these stars fall on the A', A and B overtone sequences,  whilst only a few (TW~Oph, TX~Psc and R~Scl) fall on the Mira fundamental-mode sequence C, despite the fact that these smaller-amplitude carbon stars are actually classified as semi-regulars. 

We note that masses should increase along each sequence, as predicted by theory \citep[e.g., Fig.~8 of ][]{wood90}. This is indeed the case, with our observed sample, with the exception of W~Ori (${\mathcal{M}=1.5\mathcal{M}_{\odot}}$)  and T~Cet (${\mathcal{M}=3\mathcal{M}_{\odot}}$) along sequence B. Taking into account the large uncertainty of its bolometric magnitude ${M_\mathrm{bol}=-5(1)}$, derived	 from its absolute luminosity (Table~\ref{TabFinal}), the location of the C-rich star W~Ori on sequence B is rather uncertain. Using instead ${M_\mathrm{bol}=-4}$ eliminates the problem, since it moves W~Ori to its right position along sequence C, where the three other carbon stars of our sample are located.
 
\section{Conclusion} \label{SecConcl}

We present new determinations of the angular diameter of a set of ten O-rich giants, two supergiants, and four C-rich giants, observed in the $K$-band (${\mathscr{R}=1500}$) during several runs of a few nights,  distributed over two years, using the VLTI/AMBER facility. They are obtained from the fit of synthetic SPectro-Interferometric (SPI) visibility and triple product on the true data. The synthetic SPI observables are derived by using CLV profile calculated from \textsc{marcs} model atmospheres. 

We show that the results are moderately impacted ($<$~1\% in angular diameter) by the variation of the model input parameters T$_\mathrm{eff}$, $\log g$, and $\xi_\mathrm{turb}$. During the observing period, using configurations covering different baseline angles, we find no significant variation of the angular diameter, except for TX~Psc, a result which needs to be confirmed with complementary observations. 
For the eight targets previously measured by LBI in the same spectral band, our new angular-diameter values are in good agreement with those of the literature. Except for TX~Psc, the relative deviations between our values and those of the literature are less than 5\%, which validates our method. For TX~Psc, a substantial temporal variation of the angular diameter, suspected to be correlated with the visual magnitude, could be invoked to account for the larger discrepancy.
For the eight other targets, our values are first determinations, since no angular-diameter measurement have been published yet for these stars.

These angular diameters are used to place the stars in the Hertzsprung-Russell Diagram (HRD) and to derive their masses. For stars with a known technetium content, we confront their location in the HRD to the prediction
that s-process nucleosynthesis producing technetium operates in thermally-pulsing AGB (TP-AGB) stars.
The two Tc-rich stars ($o_{1}$~Ori and TX~Psc) indeed fall along the TP-AGB, as expected. But the low-mass carbon-rich star W~Ori, despite being located close to the top of the low-mass TP-AGB, has been flagged as devoid of Tc, which, if confirmed, would put interesting constraints on the s-process in low-mass carbon stars.

Finally, we compute the pulsation constant for the pulsating stars with available periods of variation. Their location along the pulsation sequences in the Period -- Luminosity diagram confirms the mass dependency predicted by the theory, except for W~Ori and T~Cet.

Those results, based on measurements of visibilities and triple products,  illustrate the several ways to include LBI observations in the general investigation process in the field of stellar astrophysics.


\section*{Acknowledgments}
   The authors thank the ESO-Paranal VLTI team for supporting their AMBER observations, especially the night astronomers A. M\'erand, G. Montagnier, F. Patru, J.-B. Le~Bouquin, S. Rengaswamy, and W.J. de~Wit, the VLTI group coordinator S. Brillant, and the telescope and instrument operators A. and J. Cortes, A. Pino, C. Herrera, D. Castex, S. Cerda, and C. Cid. 
  A.J. is grateful to T. Masseron for his ongoing support on the use of the \textsc{marcs} code. The authors also thank the 
Programme National de Physique Stellaire (PNPS) for supporting part of this collaborative research. S.S. was partly supported by the Austrian Science Fund through FWF project P19503-N16; A.C. was supported by F.R.S.-FNRS (Belgium; grant 2.4513.11); E.P. is supported by PRODEX; K.E. gratefully acknowledges support from the Swedish Research Council. This study used the SIMBAD and VIZIER databases
at the CDS, Strasbourg (France), and NASAs ADS bibliographic services. 



\label{lastpage}

\setcounter{table}{0}

\onecolumn

\begin{landscape}
   \begin{center}
 \begin{longtable}{lcccccccccccccc}
\caption{Relevant observational parameters of the observed sample of science targets.}
\label{TabObsSci} \\
 \hline
 Name & Spec. type & $\varpi_\mathrm{Hip}$$^{a}$ & $m_\mathrm{K}$$^{b}$ & $m_{V}$$^{c}$ & $A_{V}$$^{d}$ & $(V-K)_{0}$ & Compo.$^{e}$ & Sep.$^{e}$ & PA$^{e}$ & $\Delta V$$^{e}$ & Var. type$^{c}$ & Period$^{c}$ & Tc$^{g}$ & Calib$^{h}$ \\ 
  &  & (mas) & & & & & & (\arcsec) & (\degr) &  & & (days) & & \\ 
  \hline
$\alpha$~Car & F0II$^{8}$ & 10.6(6) & -1.3(3) & -0.62(5) & 0.07(15) & 0.6(3) & - & - & - & - & - & - & unkn. & $\eta$~Col \\
$\beta$~Cet & K0III$^{13}$ & 33.9(2) &  -0.3(4) & 1.96-2.11 & 0.03(19) & 2.3(4) & - & - & - & - & SRB$^{17}$ & 37(4)$^{17}$ & unkn. & $\eta$~Cet \\
$\alpha$~TrA & K2II$^{14}$ & 8.4(2) &  -1.2(1) & 1.91(5) & 0.08(15) & 3.1(2) & - & - & - & - & - & - & unkn. & $\varepsilon$~TrA \\
$\alpha$~Hya & K3II-III$^{2}$ & 18.1(2) &  -1.1(2) & 1.93-2.01 & 0.03(14) & 3.1(3) & AB & 283 & 153 & 8 & Susp. & - & unkn. & $\lambda$~Hya \\
& & & &  & & & AC & 210 & 90 & - & - & - & unkn. & \\
$\zeta$~Ara & K3III$^{3}$ & 6.7(2) & -0.6(2) & 3.12(5) & 0.07(15) & 3.7(3) & - & - & - & - & -  & - & unkn. & $\varepsilon$~TrA/o~Sgr\\
$\delta$~Oph & M0.5III$^{2}$ & 19.1(2) & -1.2(2) & 2.72-2.75 & 0.03(15) & 3.9(3) & AB & 66 & 294 & 9 & Susp. & - & unkn. & $\gamma$~Lib/$\varepsilon$~TrA \\
$\gamma$~Hyi & M2III$^{1}$ & 15.2(1) & -1.0(4) & 3.32-3.38 & 0.04(16) & 4.2(5) & - & - & - & - & SRB & - & unkn. & $\alpha$~Ret \\
$o_{1}$~Ori & M3III$^{5}$ & 5.0(7) & -0.7(2) & 4.65-4.88 & 0.16(17) & 5.2(2) & AB$^{7}$ & - & - & 11.7 & SRB & 30(1) & yes$^{6,11}$ & HR~2411 \\
$\sigma$~Lib & M3.5III$^{4}$ & 11.3(3) &  -1.4(2) & 3.20-3.46 & 0.03(14) & 4.6(3) & - & - & - & - & SRB & 20(1) & no$^{6,11}$ & 51~Hya/$\varepsilon$~TrA \\
$\gamma$~Ret & M4III$^{3}$ & 7.0(1) & -0.5(3) & 4.42-4.64 & 0.08(15) & 4.9(4) & AB & 0.2 & - & - & SR & 25(1) & unkn. & $\alpha$~Ret \\
CE~Tau & M2Iab-b$^{3}$ & 1.8(3) & -0.9(2) & 4.23-4.54 & 0.29(19) & 5.0(3) & - & - &  - & - & SRC & 165(1) & doubt.$^{6}$ & $\phi_{2}$~Ori \\
T~Cet & M5.5Ib/II$^{8}$ & 3.7(5) & -0.8(3) & 4.96-6.9 & 0.08(19) & 6.4(3) & - & - & - & - & SRC & 159.3(1) & prob.$^{6}$ & $\iota$~Eri/$\gamma$~Scl \\     
TX~Psc & C7,2(N0)(Tc)$^{10}$ & 3.6(4) & -0.5(3) & 4.79-5.20 & 0.11(10) & 5.4(3) & - & - & - & - & LB & 220(1)$^{9}$ & yes$^{6,16}$ & $\theta$~Psc \\
W~Ori & C5,4(N5)$^{10}$ & 2.6(10)$^{15}$ & -0.5(4) & 5.5-6.9 & 0.11(15) & 6.4(5) & - & - & - & - & SRB & 212(1) & no$^{12,16}$ & $\phi_{2}$~Ori/HR~2113\\
R~Scl & C6,5ea(Np)$^{10}$ & 2.1(15)$^{15}$ & -0.1(1) & 9.1-12.9 & 0.11(10) & 6.6(2) & AB$^{f}$ & 10 & 234 & 12 & SRB & 370(1) & unkn. & $\iota$~Eri \\
TW~Oph & C5,5(Nb)$^{10}$ & 3.7(12)$^{15}$ & 0.5(4) & 11.6-13.8 & 0.10(16) & 7.0(4) &  - & - & - & - & SRB & 185(1) & unkn. & o~Sgr/$\gamma$~Lib \\
 \hline
\end{longtable}
  \end{center}
$^{1}$\citet{landi66}; $^{2}$\citet{morgan73}; $^{3}$\citet{houk75}; $^{4}$\citet{houk78}; $^{5}$\citet{smith85}; $^{6}$\citet{little87}; $^{7}$\citet{ake88}; $^{8}$\citet{houk88}; $^{9}$\citet{wasatonic97}; $^{10}$\citet{kholopov98}; $^{11}$\citet{lebzelter99}; $^{12}$\citet{abia01}; $^{13}$\citet{montes01}; $^{14}$\citet{borde02}; $^{15}$\citet{knapp03}; $^{16}$\citet{lebzelter03}; $^{17}$\citet{otero06}
\\ \\
$^{a}$unless quoted, from the New HIPPARCOS Astrometric Catalogue \citep{vleeuwen07}\\
$^{b}$from the 2MASS Catalogue \citep{skrutskie06}\\
$^{c}$magnitude variations, variability type, and period of variability taken (unless quoted) from the AAVSO-VSX Database \citep{watson06}. ``Susp.'' stands for suspected variability\\
$^{d}$calculated thanks the numerical algorithm of \citet{hakkila97}, including the
studies of \citet{fitzgerald68}, \citet{neckel80}, \citet{berdnikov91}, \citet{arenou92}, \citet{chen98},
and \citet{drimmel01}, plus a sample of studies of high-galactic latitude clouds\\
$^{e}$multiplicity parameters from the WDS Catalogue \citep{mason01}\\
$^{f}$the A component is seen double by \citet{maercker12} with ALMA \\
$^{g}$qualitative information on the technetium content (unkn. stands for ``unknown'', doubt. for ``doubtful'', and prob. for ``probable'') \\
$^{h}$associated calibrator(s), with angular diameter given by \citet{cruzalebes13}
\end{landscape}

\onecolumn

\setcounter{table}{4}

%
	\begin{center}
   \begin{longtable}{lccccccc}
\caption{Published angular diameters (in mas) of the scientific targets. We note $\phi_\mathrm{ind}$ the angular diameter derived from indirect methods, while we note $\phi_\mathrm{LO}$ and $\phi_\mathrm{LBI}$ the limb-darkened angular diameters derived from Lunar Occultation and Long-Baseline Interferometry measurements, respectively. The values in bold are the averaged values, using weights inversely proportional to the uncertainties. When not quoted, conservative 10\% errors are adopted. ``Ref.'' stands for the bibliographical reference from which each value is taken, as listed at the end of the Table, and ``Diff.'' stands for the relative difference between our new measurement and the averaged published LBI values obtained with a similar instrumental configuration. Our values are included in the $\phi_\mathrm{LBI}$ column under Ref.~(48).}
\label{TabPubDiamSci} \\ 
  \hline 
 Name & $\phi_\mathrm{ind}$  &  Ref.  & $\phi_\mathrm{LO}$ & Ref. & $\phi_\mathrm{LBI}$  &  Ref.  &  Diff.   \\
   \hline \hline
 \endfirsthead
  \caption{continued.}\\
  \hline 
 Name & $\phi_\mathrm{ind}$  &  Ref.  & $\phi_\mathrm{LO}$ & Ref. & $\phi_\mathrm{LBI}$  &  Ref.  &  Diff.   \\
\hline \hline
  \endhead
	\hline
  \endfoot
	
 $\alpha$~Car &  5.9(4)       & 16 &  &  & 6.6(8)       & 5  &  \\
              &  6.0(7)       & 3  &  &  & 6.86(41)     & 2  &  \\
              &  6.5(8)       & 11 &  &  & 6.92(11)     & 48 & -0.1\% \\
              &  6.8(4)       & 13 &  &  & 6.93(15)     & 42 &  \\
              &  7.1(2)       & 9  &  &  &              &    &  \\
              &  7.22(42)     & 37 &  &  &              &    &  \\
              &  \textbf{6.7} &    &  &  & \textbf{6.9} &    & \\
\hline
 $\beta$~Cet  &  5.03(40)     & 19 &  &  &  5.29(8)     & 46 & \\
              &  5.31(6)      & 35 &  &  &  5.329(5)    & 44 & \\
              &  5.4(8)       & 16 &  &  &  5.51(25)    & 48 & +3.4\% \\
              &  5.66(39)     & 45 &  &  &              &    &  \\
              &  6.5          & 24 &  &  &              &    &  \\
              &  7.4(9)       & 3  &  &  &              &    &  \\
              &  8.0          & 20 &  &  &              &    &  \\
              &  \textbf{5.6} &    &  &  & \textbf{5.3} &    &  \\
\hline
 $\alpha$~TrA &  11.6(17)     & 16 &  &  & 9.24(2) & 48 &  \\
              &  8.98(10)     & 35 &  &  &         &    &  \\
              &  9.81(39)     & 40 &  &  &         &    &  \\
              &  15.0(18)     & 3  &  &  &         &    &  \\
              &  \textbf{9.5} &    &  &  &         &    &  \\
\hline 
 $\alpha$~Hya &  9.30(39) & 9 &  &  & 9.73(10) & 38 &  \\ 
              &  9.4(9) & 23 &  &  & 9.335(16) & 44 &  \\ 
              &  9.9(10) & 13 &  &  & 9.36(6) & 48 &  +0.3\%\\ 
              &  10.0(15) & 16 &  &  &  &  &  \\ 
              &  14.0(17) & 3 &  &  &  &  &  \\ 
              &  \textbf{10.0} &  &  &  & \textbf{9.4} &  &   \\ 
\hline
 $\zeta$~Ara  &  7.21(21) & 39 &  &  & 7.09(12) & 48 &  \\ 
              &  7.2 & 20 &  &  &  &  &  \\ 
              &  7.6(11) & 16 &  &  &  &  &  \\ 
              &  7.62(53) & 45 &  &  &  &  &  \\ 
              &  9.0 & 24 &  &  &  &  &  \\ 
              &  11.0(13) & 3 &  &  &  &  &  \\ 
              &  \textbf{7.6} &  &  &  &  &  &  \\ 
\hline
 $\delta$~Oph & 10(1)  & 21  &  &  & 9.50(50) & 30    &     \\ 
              & 10.03(10)  & 35  &  &  & 9.93(9) & 48    &  +2.1\%   \\ 
              & 10.18(20)  & 25  &  &  & 9.946(13) & 44    &     \\ 
              & 10.22(71)  & 45  &  &  & 10.47(12) & 38    &     \\ 
              & 10.23(31)  & 40  &  &  &  &  &     \\ 
              & 11.6(17)  & 16  &  &  &  &  &     \\ 
              & 11.0  & 24  &  &  &  &  &     \\ 
              & 13  & 1  &  &  &  &  &     \\ 
              & 13.0(16)  & 3  &  &  &  &  &     \\ 
              & 26(7)  & 8  &  &  &  &  &     \\ 
              & \textbf{10.4}  &   &  &  & \textbf{10.0} &  &     \\ 
\hline
 $\gamma$~Hyi & 9.5  & 33  &  &  & 8.79(9)  &   48  &      \\ 
              & 9.8(15)  & 16  &  &  &    &     &      \\ 
              & 10.0(12)  & 3  &  &  &    &     &      \\ 
              & \textbf{9.7}  &   &  &  &    &     &      \\ 
\hline
$o_{1}$~Ori  & 7.1(21) & 3  &  &  & 9.78(10)    & 48    &      \\
\hline 
 $\sigma$~Lib & 11.0(13)      & 3     &      &  & 11.33(10)     & 48    &       \\
              & 12.05(83)      & 45     &      &  &      &     &       \\
              & 12.5      & 20     &      &  &      &     &       \\
              & 13.0      & 24     &      &  &      &     &       \\
              & \textbf{12.1}      &      &      &  &      &     &       \\
\hline
$\gamma$~Ret & 7.5(2)      & 32  &       &     & 7.44(2)     & 48    &       \\
             & 8.0      & 33  &       &     &      &    &       \\
             & 11.0(33)      & 16  &       &     &      &    &       \\
             & \textbf{7.8}      &   &       &     &      &    &       \\
\hline 
 CE~Tau       & 9.4(11)       & 3  & 9.1(8)        & 15  & 9.3(5)        & 34  &       \\ 
              & 13            & 1  & 10.9(10)      & 14  & 9.83(7)       & 30  &       \\ 
              & 13.0(20)      & 16 & 17(1)         & 18  & 9.97(8)       & 48  &   +3.7\%    \\ 
              &               &    &               &     & 10.68(21)     & 27  &       \\ 
              & \textbf{11.5} &    & \textbf{12.1} &     & \textbf{10.0} &     &       \\ 
\hline
 T~Cet        & 13.1(39)      & 16 &   &   & 9.70(8) & 48 &       \\ 
              & 14.5          & 43 &   &   &         &    &       \\ 
              & \textbf{14.1} &    &   &   &         &    &       \\ 
\hline
 TX~Psc       & 6.2          & 12 & 8.40(5)       & 29 & 10.23(36)     & 48    & -10.6\%      \\
              & 9.5(5)       & 11 & 8.9(10)       & 4  & 11.2(10)      & 28    &       \\
              &              &    & 9.31(75)      & 10 & 11.44(30)     & 31    &       \\
              &              &    & 10(3)         & 6  &               &     &       \\
              &              &    & 10.2(25)      & 7  &               &     &       \\
              & \textbf{8.0} &    & \textbf{8.5}  &    & \textbf{10.9} &     &       \\
\hline 
 W~Ori        &  &  &  &  & 9.63(4)      & 48 & -2.8\% \\ 
              &  &  &  &  & 9.91(60)     & 31 &  \\ 
              &  &  &  &  & \textbf{9.7} &  &  \\ 
\hline
 R~Scl        & 12.2     & 22      &       &     & 10.06(5)     & 48 & -1.4\%      \\  
              & 12.0     & 26      &       &     & 10.2(5)     & 47 &       \\  
              & 12.1     & 41      &       &     &      &  &       \\  
              & 12.75(98)     & 36      &       &     &      &  &       \\  
              & \textbf{12.3}     &       &       &     & \textbf{10.1}     &  &       \\  
\hline
 TW~Oph        &       &       & 10.4(5)      & 17    & 9.46(30)     & 48 &      \\  
\hline
 \end{longtable}
 \end{center}
(1)~\citet{hertzsprung22}; (2)~\citet{hanbury67}; (3)~\citet{wesselink72}; 
(4)~\citet{devegt74}; (5)~\citet{hanbury74}; (6)~\citet{morbey74}; 
(7)~\citet{dunham75}; (8)~\citet{currie76}; (9)~\citet{blackwell77}; 
(10)~\citet{ridgway77}; (11)~\citet{barnes78}; (12)~\citet{scargle79}; 
(13)~\citet{blackwell80}; (14)~\citet{white80}; (15)~\citet{beavers82}; 
(16)~\citet{ochsenbein82}; (17)~\citet{ridgway82}; (18)~\citet{white82}; 
(19)~\citet{eriksson83}; (20)~\citet{johnson83}; (21)~\citet{leggett86}; 
(22)~\citet{rowan86}; (23)~\citet{bell89}; (24)~\citet{slee89}; 
(25)~\citet{blackwell90}; (26)~\citet{judge91}; (27)~\citet{quirrenbach93}; 
(28)~\citet{quirrenbach94a}; (29)~\citet{richichi95}; (30)~\citet{dyck96a}; 
(31)~\citet{dyck96b}; (32)~\citet{bedding97}; (33)~\citet{dumm98}; 
(34)~\citet{dyck98}; (35)~\citet{cohen99}; (36)~\citet{yudin02}; 
(37)~\citet{decin03}; (38)~\citet{mozurkewich03}; (39)~\citet{ohnaka05};  
(40)~\citet{engelke06}; (41)~\citet{dehaes07}; (42)~\citet{domiciano08}; 
(43)~\citet{ramstedt09}; (44)~\citet{richichi09}; (45)~\citet{lafrasse10}; 
(46)~\citet{berio11}; (47)~\citet{sacuto11a}; (48)~present work

\begin{landscape}
  \begin{center}
   \begin{longtable}{lccccccc}
	  \caption{\label{TabFinal} Final values of the stellar fundamental parameters (top part:~science targets; bottom part:~calibrators).} \\
  \hline 
Name & $\mathcal{R}_\mathrm{obs}$/$\mathcal{R}_{\odot}$$^{a}$ & $\log T_\mathrm{eff}$$^{b}$ &  $\log \mathcal{L}_\mathrm{obs}$/$\mathcal{L}_{\odot}$$^{c}$ & $M_\mathrm{bol}$$^{d}$ &  $\mathcal{M}$/$\mathcal{M}_{\odot}$$^{e}$ & $\log g_\mathrm{obs}$$^{f}$ & $Q$$^{g}$ \\ 
\hline
$\alpha$~Car & 71(4) & 3.845(6) & 4.03(5) & -5.34(13) & 8.0(3) & 1.64(5) & - \\
$\beta$~Cet & 17.5(9) & 3.668(9) & 2.21(6) & -0.54(14) & 3.0(3) & 2.43(6) & 0.879(121)\\
$\alpha$~TrA & 119(2) & 3.638(10) & 3.66(4) & -4.41(11) & 7-8 & 1.16(3) & - \\
$\alpha$~Hya & 55.7(7) & 3.633(10) & 2.98(4) & -2.71(10) & 4-5 & 1.60(5) & - \\
$\zeta$~Ara & 114(4) & 3.628(10) & 3.58(5) & -4.21(12) & 7-8 & 1.20(4) & - \\
$\delta$~Oph & 56.0(7) & 3.562(12) & 2.70(5) & -2.01(12) & 1.0(3) & 0.93(12) & - \\
$\gamma$~Hyi & 62(1) & 3.544(12) & 2.71(5) & -2.05(13) & 1.0(3) & 0.84(12) & - \\
$o_{1}$~Ori & 214(29) & 3.538(13) & 3.76(13) & -4.65(31) & 3-4 & 0.32(13) & 0.019(4) \\
$\sigma$~Lib & 108(3) & 3.538(13) & 3.17(5) & -3.18(14) & 1.5-2 & 0.61(7) & 0.024(2) \\
$\gamma$~Ret & 115(2) & 3.538(13) & 3.23(5) & -3.33(13) & 1.5-2 & 0.55(6) & 0.027(2) \\
CE~Tau & 601(83) & 3.531(13) & 4.63(13)& -6.83(32) & 8.0(3) & -0.21(12) & 0.033(7) \\
T~Cet & 275(34) & 3.531(13) & 3.91(12) & -5.03(30) & 3.0(3) & 0.01(11) & 0.059(11) \\     
TX~Psc& 293(66)& 3.512(13) & 3.90$\left(^{21}_{31}\right)$ & -5.02(61) & 1.8(3) & -0.30(21) & 0.056(17) \\
W~Ori & 406(185) & 3.415(17) & 3.83$\left(^{51}_{37}\right)$ & -4.85(103) & 1-2 & -0.60$\left(^{42}_{63}\right)$ & 0.032$\left(^{30}_{20}\right)$ \\
R~Scl & 513(721) & 3.415(17) & 4.04$\left(^{117}_{55}\right)$ & -5.35(201) & 2-3 & -0.59(75) & 0.050$\left(^{76}_{44}\right)$ \\
TW~Oph & 278(102) & 3.415(17) & 3.50$\left(^{42}_{34}\right)$ & -4.02(89) & 1.0(3) & -0.46$\left(^{38}_{47}\right)$ & 0.040$\left(^{32}_{21}\right)$ \\
 \hline
$\alpha$~Ret & 13.5(3) & 3.679(9) & 1.93(4) & -0.10(11) & 2.5-3 & 2.61(5) & - \\
$\varphi_{2}$~Ori & 8.8(1) & 3.669(9) & 1.52(4) & 0.95(10) & 1-2 & 2.73$\left(^{13}_{19}\right)$ & - \\
$\eta$~Col & 37.1(12) & 3.668(9) & 2.77(5) & -2.17(12) & 5.0(3) & 2.00(4) & - \\
$\lambda$~Hya & 9.7(8) & 3.668(9) & 1.60(8) & 0.74(20) & 1-2 & 2.64$\left(^{20}_{25}\right)$ & - \\
$\gamma$~Lib & 12.4(6) & 3.668(9) & 1.81(6) & 0.21(14) & 2.0(3) & 2.55(7) & - \\
$o$~Sgr & 11.7(9) & 3.668(9) & 1.76(7) & 0.33(19) & 2.0(3) & 2.60(8) & - \\
$\iota$~Eri & 11.7(10) & 3.663(9) & 1.74(8) & 0.40(21) & 1-2 & 2.48$\left(^{20}_{25}\right)$ & - \\
$\theta$~Psc & 10.4(7) & 3.662(9) & 1.63(7) & 0.67(17) & 1-1.5 & 2.50(14) & - \\
$\gamma$~Scl & 12.3(1) & 3.654(10) & 1.75(4) & 0.37(10) & 1-1.5 & 2.35(10) & - \\
HR~2113 & 35(4) & 3.647(10) & 2.62(10) & -1.82(24) & 3-4 & 1.90(11) & - \\
$\varepsilon$~TrA & 16.2(2) & 3.647(10) & 1.96(4) & -0.16(10) & 1-2 & 2.20$\left(^{14}_{19}\right)$ & - \\
$\eta$~Cet & 13.6(1) & 3.647(10) & 1.81(4) & 0.22(10) & 1-1.5 & 2.26(9) & - \\
HR~3282 & 78(6) & 3.636(10) & 3.28(8) & -3.45(20) & 6-7 & 1.47(8) & - \\
HR~2411 & 22.7(10) & 3.629(10) & 2.18(6) & -0.72(14) & 1-2 & 1.90$\left(^{16}_{21}\right)$ & - \\     
51~Hya & 11.6(6) & 3.629(10) & 1.60(6) & 0.74(16) & 1.0(3) & 2.30(12) & - \\
\hline
  \end{longtable}
\end{center}
$^{a}$Rosseland radius derived from the angular diameter $\phi$ and the parallax $\varpi$\\
$^{b}$effective temperature of the model (in K)\\
$^{c}$empirical stellar luminosity, derived from Eq.~(\ref{logL})\\
$^{d}$bolometric magnitude derived from the stellar luminosity\\
$^{e}$stellar mass derived from the position along the evolutionary track in the HRD. If two tracks with different masses pass through the star location, two possible mass values are listed\\
$^{f}$surface gravity, derived from Eq.~(\ref{logg})\\
$^{g}$pulsation constant (in days), derived from Eq.~(\ref{PulsCst})
\end{landscape}
  

\twocolumn

\end{document}